\documentclass[fleqn,usenatbib]{mnras}

\usepackage{newtxtext,newtxmath}

\usepackage[T1]{fontenc}

\DeclareRobustCommand{\VAN}[3]{#2}
\let\VANthebibliography\thebibliography
\def\thebibliography{\DeclareRobustCommand{\VAN}[3]{##3}\VANthebibliography}

\usepackage{graphicx}	
\usepackage{amsmath}	
\usepackage{epsfig,longtable,color}
\usepackage[english]{babel}

\newcommand{\myemail}{h.stiele@fz-juelich.de}

\newcommand{\mr}{\mathrm}

\def\eg{e.\,g.}                                      

\def\nus{\textit{NuSTAR}}
\def\nice{\textit{NICER}}

\def\gx339{GX\,339--4}
\def\h1743{H\,1743--322}

\def\sw19{Swift\,J1910.2--0546}

\def\ga{\mathrel{\mathchoice {\vcenter{\offinterlineskip\halign{\hfil
$\displaystyle##$\hfil\cr>\cr\sim\cr}}}
{\vcenter{\offinterlineskip\halign{\hfil$\textstyle##$\hfil\cr
>\cr\sim\cr}}}
{\vcenter{\offinterlineskip\halign{\hfil$\scriptstyle##$\hfil\cr
>\cr\sim\cr}}}
{\vcenter{\offinterlineskip\halign{\hfil$\scriptscriptstyle##$\hfil\cr
>\cr\sim\cr}}}}}                                            

\title[QPOs in the 2021 outburst of \gx339]{A journey from the hard to the soft state: How do QPOs evolve in the 2021 outburst of \gx339?}

\author[Stiele \& Kong]{
H. Stiele,$^{1,2}$\thanks{E-mail: \myemail}
A. K. H. Kong,$^{2}$
\\
$^{1}$J\"ulich Supercomputing Centre, Forschungszentrum J\"ulich, Wilhelm-Johnen-Stra\ss e, 52428 J\"ulich, Germany\\
$^{2}$Institute of Astronomy, National Tsing Hua University, No.~101 Sect.~2 Kuang-Fu Road,  30013, Hsinchu, Taiwan
}

\date{Accepted 2023 March 27. Received 2023 March 27; in original form 2022 November 14}

\pubyear{2023}

\begin{document}
\label{firstpage}
\pagerange{\pageref{firstpage}--\pageref{lastpage}}
\maketitle

\begin{abstract}
We investigated the snapshots of five \nice\ observations of the black hole transient \gx339\ when the source transited from the hard state into the soft state during its outburst in 2021. 
In this paper, we focused our study on the evolution of quasi-periodic oscillations (QPOs) and noise components using power-density spectra. In addition, we derived hardness ratios comparing count rates above and below 2 keV. 
The evolution from the hard to the soft state was a somewhat erratic process showing several transitions between states that are dominated by top-flat noise and can show type-C QPOs; those that are dominated by red noise and can show type-B QPOs. From the parameters that we studied, we only found a strong correlation between the hardness ratio and the type of QPO observed.  This implies that the appearance of type-B QPOs is related to a change in the accretion geometry of the system that also reflects in altered spectral properties.    
We also observed that the type-B QPO forms from or disintegrates into a broad peaked feature when the source comes out of or goes to the hard-intermediate state, respectively. This implies some strong decoherence in the process that creates this feature. 
\end{abstract}




\section{Introduction}

\gx339\ can be considered the textbook example of a transient low-mass black hole X-ray binary. A detailed introduction to spectral and timing studies of black hole X-ray binaries in general and to \gx339\ in particular can be found \eg\ in \citet{2011MNRAS.418.2292M,2015ApJ...808..122F,2017ApJ...844....8S}. 

In this paper, we will focus on the transition from the hard intermediate (HIMS) to the soft intermediate state (SIMS) as observed by \nice\ monitoring observations of the 2021 outburst of \gx339. For a definition of these states and their properties see \citet{2006csxs.book..157M,2010LNP...794...53B}. During the HIMS, type-C quasiperiodic oscillations (QPOs) can be observed, while the SIMS is characterised by the presence of type-B QPOs. An overview of the different types of QPOs can be found in \citet{1999ApJ...526L..33W,2005ApJ...629..403C,2011BASI...39..409B,2014SSRv..183...43B,2019NewAR..8501524I}. While type-C QPOs are believed to be caused by Lense-Thiering precession of a hot, geometrically thin, optically thick inner accretion flow located in the vicinity of the accreting black hole \citep{1998ApJ...492L..59S,2009MNRAS.397L.101I}, the nature and cause of type-B QPOs is less clear. From studies of \gx339\ based on RXTE data \citep[see \eg\ ][]{2011MNRAS.418.2292M} it is known that the transition from the HIMS to the SIMS is not smooth and that it shows excursions back to the HIMS, however these studies provide a coverage of about 1 observation per day and there is only one case where a transition from the HIMS to the SIMS was observed in observations that are half a day apart. With the \nice\ observations studied here we are able to investigate changes on the time scales of hours.  

A comprehensive study of the spectral and temporal variability properties of the \nice\ monitoring observations of the 2021 outburst of \gx339\ will be presented in a future paper.

\gx339\ is known to show outbursts quite frequently \citep[\eg\ ][]{2002MNRAS.329..588K}. Renewed activity of the source was detected on 2021 January 25 \citep{2021ATel14351....1P}. The outburst is studied with many different X-ray instruments, including \nus\ \citep{2021ATel14352....1G,2021ATel14484....1M}, \nice\ \citep{2021ATel14384....1W,2021ATel14490....1W}, INTEGRAL \citep{2021ATel14354....1S,2021ATel14440....1F}, AstroSat \citep{2021ATel14400....1H,2021ATel14455....1B}, and Insight-HMXT \citep{2021ATel14504....1L}. A study of the spectral properties during the hard-to-soft state transition of this outburst as seen with Insight-HMXT was reported by \citet{2022arXiv221109543L}. \citet{2023MNRAS.519.1336P} used a model with two Comptonising regions to investigate the rms and lag spectra as well as the time-averaged spectra of the type-B QPOs in this outburst based on \nice\ and \textit{AstroSat} data.
 
\begin{figure}
\centering
\resizebox{\hsize}{!}{\includegraphics[clip,angle=0]{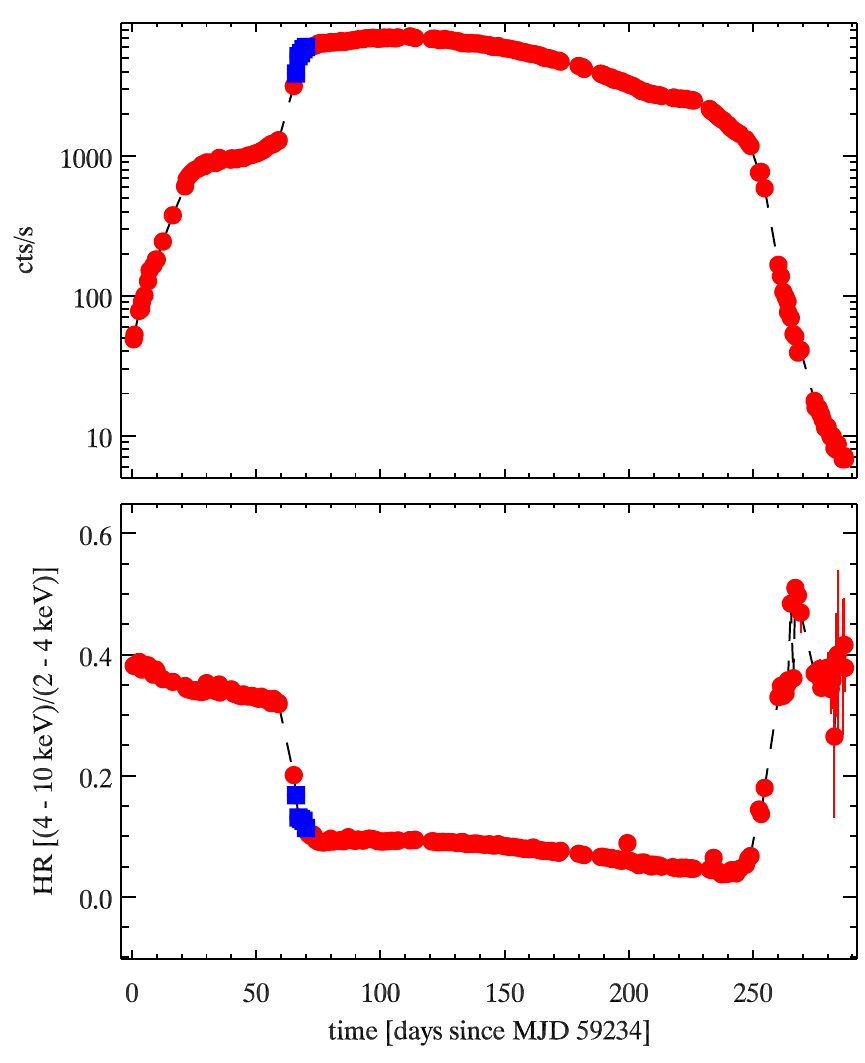}}
\caption{Light curve (upper panel) and evolution of hardness ratio (lower panel) of the 2021 outburst, based on \nice\ count rates. Each data point represents one observation. The observations that are investigated in more detail in this study are marked by (blue) squares. T=0 corresponds to January 20th 2021 00:00:00.000 UTC.}
\label{Fig:lc}
\end{figure}

\begin{figure}
\centering
\resizebox{\hsize}{!}{\includegraphics[clip,angle=0]{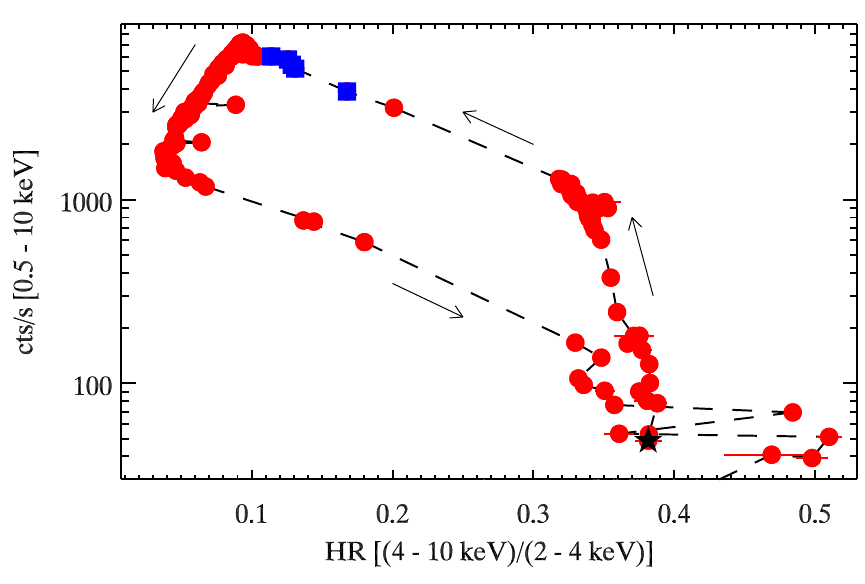}}
\caption{Hardness-intensity diagram of the 2021 outburst for \nice\ observations with more than 20 counts/s in the 0.5 -- 10 keV range. Each data point represents one observation. The observations that are investigated in more detail in this study are marked by (blue) squares. The first observation, taken on January 20th 2021, is marked by a (black) star.}
\label{Fig:hid}
\end{figure}

\section[]{Observation and data analysis}
\label{Sec:obs}

\subsection{\nice}
\nice\ \citep[The Neutron star Interior Composition Explorer;][]{2012SPIE.8443E..13G} followed the 2021 outburst of \gx339\ from January 20 until November 03. For each observation we obtain count rates in four energy bands: total (0.5 -- 10 keV), soft (0.5 -- 2 keV), medium (2 -- 4 keV), and hard (4 -- 10 keV). A light curve showing the evolution of \nice\ count rate during the 2021 outburst, the evolution of the hardness ratio -- obtained by dividing the count rates in the hard and medium bands -- and the hardness-intensity diagram of this outburst are shown in Figs.\,\ref{Fig:lc} and \ref{Fig:hid}. Here we focus our analysis on five observations (indicated as (blue) squares in Figs.\,\ref{Fig:lc} and \ref{Fig:hid}) that cover the transition from the HIMS to the SIMS and into the soft state. These observations were taken between March 27 and 31, and the Obs.~ids.\ are: 4133010104, 4133010105, 4133010106, 4133010107, and 4133010108.

To study the timing properties of these observations we make use of the pre-processed event files distributed by the \nice\ data center to derive Poissonian noise subtracted, Leahy normalised power density spectra (PDS) covering the total energy range and the $6.1\times10^{-3}$ -- 50 Hz frequency range. The PDS are converted to square fractional rms normalisation \citep{1990A&A...227L..33B}. 

A \nice\ observation may contain more than one contiguous pointing at the target, due to the orbit of the International Space Station and visibility constraints. A contiguous pointing is known as a snapshot.\footnote{\url{https://heasarc.gsfc.nasa.gov/docs/nicer/data_analysis/workshops/NICER-Workshop-Analysis-Markwardt-2021.pdf}} To follow the evolution of the noise components and the QPOs with sufficient statistics we averaged the individual PDS of each snapshot (that way the gaps in the \nice\ data do not affect our analysis). In total we analyse 53 snapshots. To study the energy dependence of the PDS we used the \textsc{heasoft} task \texttt{niextract-events} to produce event files in the 0.3--2 and 2--10 keV range, respectively. We derived PDS in the same way as for the total energy range.  

For the time intervals of the averaged PDS, we determine count rates in three energy bands: total (0.5 -- 10 keV), soft (0.5 -- 2 keV), and hard (2 -- 10 keV). (Since there are fewer counts in the individual time intervals than in the total observation, we use a single 2 -- 10 keV (hard) band for the intervals, instead of the medium and hard bands used for the HRs of observations that are shown in Figs.\,\ref{Fig:lc} and \ref{Fig:hid}.) Hardness ratios (HRs) are derived by dividing the count rate obtained in the hard band by the one observed in the soft band. We also extracted light curves in the 12 -- 15 keV band, which is assumed to be free of any astronomical signals, of the whole observations to check whether they are affected by background flares. All the observations discussed here are free from background flares based on examinations of the 12 -- 15 keV light curves.

\section[]{Results}
\label{Sec:res}

\begin{figure}
	\resizebox{\hsize}{!}{\includegraphics[clip,angle=0]{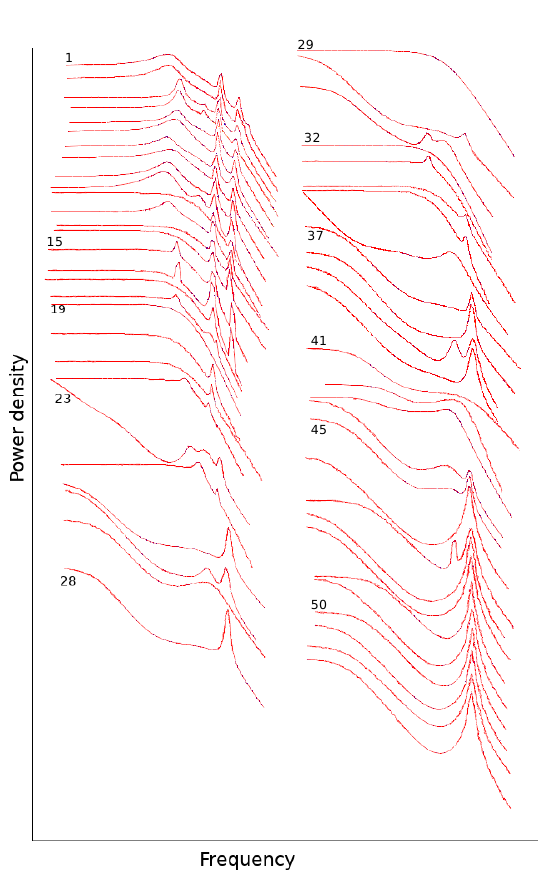}}
	\caption{Evolution of the power density spectra. Shown are the best-fits to the PDS of the individual snapshots to get an impression of the overall evolution of the different components of the PDS, especially the QPOs. The increase of the characteristic frequency of the QPO is not visible here, as we aligned the QPOs to plot the PDS more compactly. More detailed PDS, including data points, are presented in Fig.~\ref{Fig:pds1}.}
	\label{Fig:pds_evol}
\end{figure}

\begin{figure}
\resizebox{\hsize}{!}{\includegraphics[clip,angle=0]{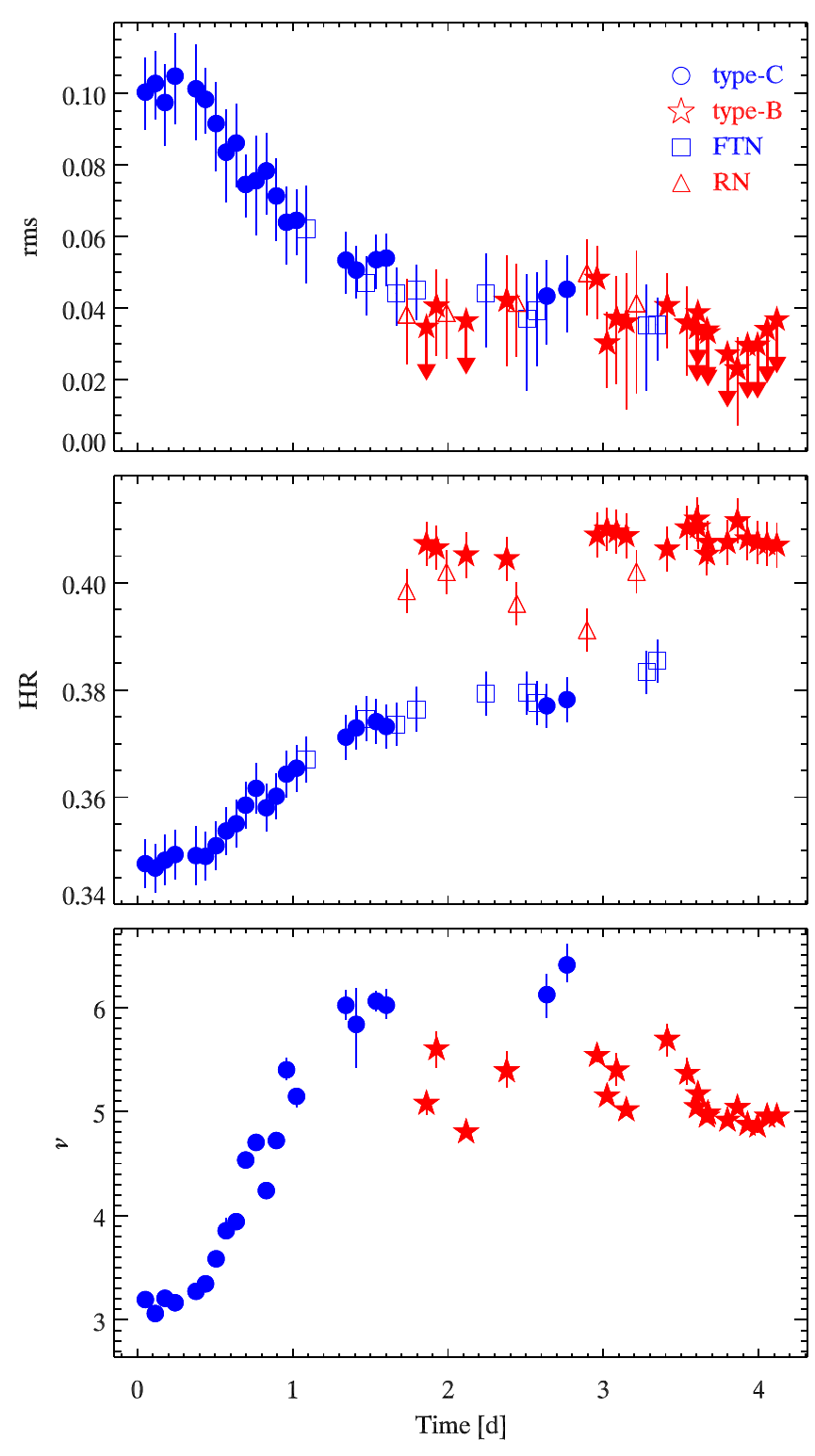}}
\caption{Evolution of the rms, hardness ratio (count rates in the (2 -- 10 keV)/(0.5 -- 2.0 keV) bands) and characteristic frequency of the fundamental QPO (from top to bottom) as \gx339\ evolves from the HIMS to the SIMS during its 2021 outburst. Filled symbols indicate detections of QPOs at more than 3$\sigma$. Arrows indicate intervals for which only upper limits on the rms value can be obtained. $T=0$ corresponds to 2021-03-27 00:00:00.000 UTC.}
\label{Fig:pds_prop_tr}
\end{figure}

\begin{figure}
\resizebox{\hsize}{!}{\includegraphics[clip,angle=0]{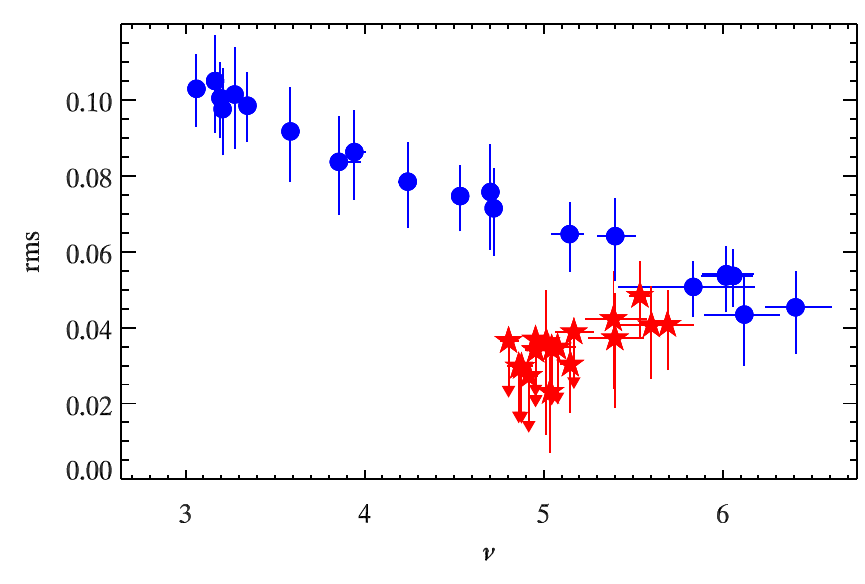}}
\caption{Shown is the rms variability versus the characteristic frequency of the fundamental QPO for type-C and type-B QPOs. Filled symbols indicate detections of QPOs at more than 3$\sigma$. The shape of the symbols have the same meaning as in Fig.~\ref{Fig:pds_prop_tr}.}
\label{Fig:rms_qpo_tr}
\end{figure}

\subsection{Power density spectra}
We fit the PDS using zero-centred Lorentzians for band-limited noise (BLN) components, and Lorentzians for QPOs. The obtained parameters for the BLN and the QPOs are given in Tables \ref{Tab:PDS}--\ref{Tab:QPOaf}. The overall evolution of the PDS and their different components can be obtained from Fig.~\ref{Fig:pds_evol}. More detailed PDS are presented in Fig.~\ref{Fig:pds1}. The evolution of the PDS during the hard-to-soft state transition is quite complicated. For the first 14 snapshots (observation 4133010104) the PDS can be well described by a BLN component, a peaked noise (PN) component and a QPO and its upper harmonic. In some of the snapshots a lower harmonic is also present. Throughout these snapshots we observe an increase of the characteristic frequency of the fundamental QPO from $\sim3$ Hz to $\sim5$ Hz. In the next two snapshots the PN component is more peaked and its Q factor is high enough to regard this feature as a QPO. The PDS of snapshot 17 shows BLN and the fundamental QPO, while the upper harmonic is no longer detected. In the next snapshot the PN is detected again, while the QPO disappears. In snapshot 19 only BLN can be detected. The next two snapshots show again the fundamental QPO together with BLN. The PDS of snapshot 22 shows BLN and PN. There also may be a peaked feature at $\sim4$ Hz, but its significance is below 2$\sigma$.  Then some dramatic change takes place as in the next snapshot the flat-top noise seen up-to-now, is replaced by red noise and several features are visible between $\sim2-6$ Hz. In snapshot 24 the flat-top noise together with the PN returns and the feature around 4 Hz also shows up again. In the next snapshot we once more observe red noise and it is the first time that a type-B QPO can be observed in PDS of the 2021 outburst of \gx339. In the then following snapshot, the type-B QPO is weaker and a subharmonic appears, before the two features merge into a broad PN component in snapshot 27. In snapshot 28 the type-B QPO and red noise are detected one more time. The next snapshot shows again BLN, before the source returns to a state where the PDS is dominated by red noise and the type-B QPO is observed anew. In the then following snapshot the type-B QPO is once more replaced by a broad feature, before \gx339\ returns to a state where the PDS show BLN for the next four snapshots. In the second of these four snapshots, the PN is detected again, while in the third and fourth the type-C QPO is found one more time. In the next snapshot the red noise returns together with a broad peaked feature. The PDS of snapshots 37 to 40 show again a type-B QPO, where in snapshot 39 also a subharmonic is present. The next snapshot only shows red noise, before \gx339\ returns once more to a state where the PDS is dominated by BLN and PN for two snapshots. The PDS of snapshot 44 shows red noise for an additional time, a type-B QPO and a broader peaked feature at a lower frequency. In the next snapshot the type-B QPO is more significant and the peaked feature weakens. The type-B QPO is present in all remaining snapshots before \gx339\ transits into the soft state. During these 8 snapshots the noise component contributes less.    

Regarding the evolution of the characteristic frequency of the fundamental QPO we find an increase from $\sim3$ Hz to $\sim6$ Hz for the type-C QPO. The type-B QPO appears at a characteristic frequency of about 5--5.5 Hz, which is quite close to the lastly seen characteristic frequency of the type-C QPO. The return of the type-C QPO happens at a characteristic frequency of 6--6.5 Hz, continuing the increase of the frequency seen before the appearance of type-B QPOs. Finally the type-B QPO is again found in the 5.5--5 Hz range, seeming to show an evolution towards lower frequencies. 

Figure~\ref{Fig:pds_prop_tr} shows the evolution of the centroid frequency of the fundamental QPO, of the HR (count rates in the (2 -- 10 keV)/(0.5 -- 2.0 keV) bands) and the rms. From this figure it is obvious that red noise and type-B QPOs appear above a HR value of 0.39, while flat-top noise and type-C QPOs are observed at lower HR values. Regarding the rms values, we do not find a clear distinction between snapshots that show red noise/type-B QPOs and those that show flat-top noise/type-C QPOs.

The correlation between the rms variability and the characteristic frequency of the fundamental QPO is shown in Fig.~\ref{Fig:rms_qpo_tr}. For type-C QPOs we observe the expected anti-correlation between the two parameters, while type-B QPOs cluster at lower rms values. 

\subsubsection{PDS in different energy bands}
We also investigate PDS in the soft (0.3--2) and hard (2--10 keV) bands. In the 14 snapshots of observation 4133010104, the QPOs are clearly present in both energy bands. In snapshots 17, 20 and 21 the fundamental QPO is more prominent in the soft band. The features between $\sim2-6$ Hz in snapshot 23, when we observe flat-top noise for the first time during this outburst, are present in both energy bands.  The type-B QPO that is detected in snapshots 25 and 26 is present in both bands, but it is more prominent in the hard band, while the subharmonic is stronger in the soft band. In snapshot 27, where we observe a broad PN component in the total band, an additional bump at the frequency of the type-B QPO is visible in the hard band. The type-B QPO in snapshots 28 and 30, and the broad feature in snapshot 31 are also detected in both bands, but again they are more obvious in the hard band, while the bump at lower frequencies (subharmonic) is clearly stronger in the soft band. For the next four snapshots \gx339\ returns to the HIMS with the PDS of the total energy band of the last two snapshots (34 and 35) showing a type-C QPO. This QPO is in the soft band, but it is not detectable in the hard band. Actually, the type-C QPO already shows up in the soft band PDS of snapshot 33, but it is too weak to be detectable in the full band PDS of this snapshot that is dominated by BLN and PN.
The broad peaked feature in snapshot 36 is visible in both energy bands. In the remaining snapshots where a type-B QPO is detected it is present in both bands, but stronger in the hard band, whereas the subharmonic, if present, is more prominent in the soft band. 

\section[]{Discussion}
\label{Sec:dis}
In this study we made use of data obtained during the 2021 outburst of \gx339\ with \nice. In particular, we investigate the evolution of the QPOs and noise components as \gx339\ evolves from the hard-intermediate state (HIMS), where type-C QPOs are observed, through the soft-intermediate state (SIMS), which is defined by the presence of type-B QPOs, into the soft state.  Thanks to the enhanced sensitivity of \nice\ compared to previous missions like RXTE and its monitoring capabilities, we are able to follow the evolution of different types of QPOs and associated noise properties on hours to minutes time scales. Remarkably, all these complex changes occur within hours.

Here we follow the development of the QPOs and noise components  over a period of four days, and study the PDS of individual snapshots. We observe an increase of the characteristic frequency of the fundamental type-C QPO from $\sim3$ Hz to $\sim6$ Hz. This evolution of the QPO frequency along the outburst has been observed in previous outbursts of \gx339\ and is a common behaviour in the evolution of canonical black hole XRBs \citep[see \eg\ ][]{2017ApJ...845..143Z,2016ApJ...828...34F,2011MNRAS.418.2292M}. We also detect the upper harmonic of the type-C QPO and its characteristic frequency increases from $\sim 6$Hz to $\sim 11$Hz. We then observe a weakening of the type-C QPO, beginning with a non-detection of its upper harmonic, and a decrease of the amplitude of the fundamental QPO itself. This weakening of the type-C QPO is intercepted by snapshots where only noise components are detected. We then observe the appearance of a red noise component at lower frequencies, while at frequencies above a few Hz, the BLN still contributes. The source then returns shortly to a state dominated by BLN noise, before the first detection of a type-B QPO during the 2021 outburst. The characteristic frequency of the type-B QPO ($\sim5$ Hz) is a little bit lower than the characteristic frequency of the last type-C QPO ($\sim6$ Hz). We then observe a ``decay'' of the type-B QPO and the appearance of a broad peaked feature at $\sim2.5$ Hz and a merging with the type-B QPO into a broad feature, indicating some strong decoherence in the process that causes this feature. After a gap between observations of $\sim10$ ks the type-B QPO is found again. This is followed by snapshots in which \gx339\ transits between states dominated by BLN and those dominated by red noise, before returning to a state dominated by flat-top noise and a reappearance of the type-C QPO at $\sim6$ Hz. The source then again goes into a state showing red noise and a type-B QPO that is followed by yet another short excursion to a state dominated by BLN and PN before \gx339\ finally settles into the SIMS showing red noise and a type-B QPO. During this part of the outburst we observe a weak red noise that contributes at all observed frequencies.   

We also investigate the energy dependence of the QPOs by studying PDS in the soft (0.3--2) and hard (2--10 keV) energy bands. We observe the type-C QPOs in the first observation that we investigated in both energy bands, as it is expected for a black hole X-ray binary being in the HIMS. We also found that the fundamental of the type-B QPO is stronger in the hard band, consistent with what has been found in previous studies \citep[see \eg\ ][]{2014MNRAS.438..341G,2020MNRAS.496.4366B}. More difficult to explain is the finding that the reappearance of the type-C QPO takes place mainly in the soft band. As the source evolves towards the soft state, the soft band should be more and more dominated by the emission of the disk. As type-C QPOs are thought to be caused due to Lense-Thirring precession of a hot inner flow in a truncated accretion disc geometry \citep{1999PhRvL..82...17S,2009MNRAS.397L.101I} one would expect to observe them in the hard band that is still dominated by the emission of the corona. Of course, the reappearance of the type-C QPO happens during excursions back to a harder state. However, one would still expect to observe them more clearly in the hard band than in the soft band. The interpretation of this finding is complicated by the change of the HR during periods in which type-B and type-C QPOs are detected, as this HR change indicates alteration in the contribution of the flux in the hard band according to \citet{2011MNRAS.418.2292M}. 
 
Regarding the mechanism that creates the QPOs, it is nowadays widely accepted that type-C QPOs can be attributed to Lense-Thirring precession of a hot inner flow in a truncated accretion disc geometry \citep{1999PhRvL..82...17S,2009MNRAS.397L.101I}. In addition to the hot inner flow, which drives the type-C QPO, and the thin accretion disc, there will be a transitional region where a pasive disc is imbedded in the outer, non-thermal corona in some way. The characteristic frequency of the type-C QPO is anti-correlated to the radial extent of the combined Comptonised emission region. The type-C QPO gets quenched when the accretion disc has moved inward far enough that the hot flow region gets too small to produce detectable type-C QPOs, while the energy spectra can still show Comptonised emission that comes from the region where the outer corona interacts with the accretion disc \citep{2023arXiv230208314K}. A temporary increase of the radius of the combined Comptonised emission region will lead to a reappearance of the type-C QPO at a frequency close to the one where it was last seen before the transition to the SIMS, in agreement with the evolution of the characteristic frequency of the type-C QPOs seen in this study.

The cause of type-B QPOs is not well understood and several different mechanisms are under debate. They can be either related to some type of instability \citep{1999ApJ...518L..95T,1999A&A...349.1003T,2001ApJ...554.1210L} or they are somehow connected to the relativistic jets \citep{2009MNRAS.396.1370F,2012MNRAS.421..468M}. 
The results shown here, clearly support the idea that the two different QPO types are caused by two distinct mechanisms. Another observational hint at the existence of two distinct mechanism is the simultaneous detection of type-B and C QPOs in the PDS of one observation of GRO\,J1655--40 in its ultraluminous state \citep{2012MNRAS.427..595M}. 
Our study shows that the transition from the HIMS to the SIMS is a gradual process that is intermitted by excursions back to the HIMS, implying that changes of QPO type is an interplay between two processes. Similar results have been found from studies of RXTE data \citep{2011MNRAS.418.2292M}, however on a much coarser time scale, compared to the scale of hours that is accessible with the dense \nice\ coverage. 

We also find that the red noise first shows up on lower frequencies (or longer time scales) while the higher frequencies still show BLN. Consistent with results of previous studies, we find that there is not much evolution in the characteristic frequency of the type-B QPO. However, we observe that during the phase of the outburst when \gx339\ shuttles between the HIMS and SIMS the type-B QPO forms from or disintegrates into a broad feature, sometimes showing a ``sub-harmonic'' peak before the source returns to or when it comes out of the HIMS. This behaviour indicates some strong decoherence in the process that causes the type-B QPO. The presence of a broad feature at frequencies where a type-B QPO forms in another observation was also observed in the ultraluminous state of GRO\,J1655--40 \citep{2012MNRAS.427..595M}. 

\citet{2022A&A...662A.118M} suggest a model based on radiative feedback between a hot electron population in the corona and soft radiation emerging from the accretion disc, due to reprocessing of the hard corona radiation. The temporal properties of this model are similar to the classical dynamic system of prey and predator, and damped oscillations identified with type-C QPOs are observed. Because of the damping, the oscillations need to be sustained by small fluctuations in the accretion rate. To quench the type-C QPOs, the electrons need to escape quickly from the corona, which will happen when the corona shrinks and the outflow becomes continuously narrower as the source evolves from the hard to the soft states. In the SIMS the outflow is narrow enough to show type-B QPOs that, according to \citet{2020A&A...640L..16K}, can be explained quantitatively as precession of the outflow. Furthermore, an increase in soft disc radiation supports the suppression of type-C QPO activity in the SIMS. In light of the results presented here, the shrinking of the corona cannot be a uni-directed process and/or there must be fluctuations in the soft disc radiation. To study this further one would need to investigate if the changes in the corona size and soft disc radiation can occur on the timescales observed here, which is beyond the scope of the present work. We observe a continuous evolution and increase of the characteristic frequency of the type-C QPOs after the first appearance of type-B QPOs. If the evolution of type-C QPOs is driven by an increase in accretion rate, this increase may in some part counteract the shrinking of the corona and hence explain the reappearance of the type-C QPOs.

Based on a spectral--timing Comptonisation model \citep{2020MNRAS.492.1399K}, \citet{2021MNRAS.501.3173G} showed that the type-B QPOs of MAXI\,J1348--630 (including their energy-dependent fractional-rms and phase-lag spectra) can be explained by the presence of two Comptonisation regions with different sizes and located at different distances from an accretion disk. \citet{2023MNRAS.519.1336P} applied a Comptonisation model with two components to the type-B QPOs in the 2021 outburst of \gx339. In these two-component Comptonisation models the bigger corona is presumably related to the base of the jet. How type-B QPOs can be produced through Comptonisation processes in a precessing jet is discussed in  \citet{2020A&A...640L..16K}. The lags seen in the type-C QPOs of GRS\,1915+105 can also be explained by a two-component corona \citep{2000ApJ...538L.137N}. \citet{2021MNRAS.503.5522K} proposed a similar model based on \citet{2020MNRAS.492.1399K} for the type-C QPOs in GRS\,1915+105 that can also explain the rms- and lag-energy spectra at different QPO frequencies. The different physical assumptions have an impact on the correlation between the inner disc radius and the corona size. In \citet{2021MNRAS.503.5522K} the corona and disc size are anti-correlated for frequencies $\ga2$ Hz and correlated for lower frequencies. The large size of the corona at high type-C QPO frequencies makes it quite challenging to find in this model an explanation for the rather fast ``oscillations'' between type-C and B QPOs that we report here. Even if the corona that is responsible for the type-C QPO shrinks in size, as is assumed in the other models mentioned above, there remains the following point. If both, type-C as well as type-B QPOs, are due to a two-component corona, then the evolution observed in our study needs to be explained by an evolution of the two Comptonisation regions that allows the system to kind of ``oscillate'' between the two states when it transits from the HIMS to the SIMS.     

We also find a clear distinction in the HR values of the snapshots that show flat-top noise/type-C QPOs and those that show red noise/type-B QPOs. This finding indicates that the transition between these two states is related to a change in the accretion geometry that shows up in a transition of the spectral properties. While we observe an overall decrease of the (4 -- 10 keV)/(2 -- 4 keV) HR with time during the observations studied here (see Fig.\,\ref{Fig:hid}), the (2 -- 10 keV)/(0.5 -- 2 keV) HR increases (see Fig.\,\ref{Fig:pds_prop_tr}). This means that the (relative) count rate in the 2 -- 4 keV band increases compared to the count rates in the 4 -- 10 keV and 0.5 -- 2 keV bands. The increase in the (2 -- 10 keV)/(0.5 -- 2 keV) HR can be interpreted as a stronger contribution of the blackbody emission of the disc in the 2 -- 4 keV band (and a simultaneous smaller contribution in the 0.5 -- 2 keV band, since the maximum of disc blackbody emission moves towards higher energies with increasing disc temperature). According to \citet{2011MNRAS.418.2292M} the decrease of the (4 -- 10 keV)/(2 -- 4 keV) HR is due to changes in the contribution of flux in the hard band.

The behaviour observed in this study reveals that the transition between type-B and type-C QPOs is a rather slow process taking place on the scale of hours. This slow evolution is not in conflict with results of \citet{2003A&A...412..235N} on the fast evolution of the characteristic frequency of the QPO on time scales of 10 seconds, as after its appearance the QPO has been present during the whole remaining part of the RXTE observation. Based on RXTE data, \citet{2011MNRAS.418.2292M} showed that transitions between type-A and type-B QPOs can occur on time scales of a few seconds in \gx339.   

\section[]{Conclusions}
Investigating PDS derived from \nice\ data, we found that during the hard-to-soft state transition of its 2021 outburst \gx339\ altered several times on the timescales of hours between the HIMS that shows type-C QPOs and the SIMS where type-B QPOs are observed.  The rather fast ``oscillations'' between type-C and type-B QPOs and the associated change of spectral properties seen in the hardness ratio, clearly support the idea that the two different QPO types are caused by two distinct physical mechanisms. Independent of its exact composition, a successful model of the origin of QPOs should be able to reproduce this ``oscillating'' behaviour between the two states and the related spectral and timing properties.

\section*{Acknowledgements}

We thank the anonymous referee for thoughtful comments that helped us to improve the discussion of our results.
This project is supported by the Ministry of Science and Technology of
the Republic of China (Taiwan) through grants 110-2628-M-007-005 and 111-2112-M-007-020. HS acknowledges support from the ``Big Bang to Big Data'' (B3D) project, the NRW cluster for data-intensive radio astronomy, funded by the state of North Rhine-Westphalia as part of the \textit{Profiling 2020} programme. 

\section*{Data Availability}

This research has made use of publicly available data obtained through the High Energy Astrophysics Science Archive Research Center Online Service (\url{https://heasarc.gsfc.nasa.gov/docs/archive.html}), provided by the NASA/Goddard Space Flight Center.



\bibliographystyle{mnras}
\bibliography{} 

\begin{thebibliography}{}
\makeatletter
\relax
\def\mn@urlcharsother{\let\do\@makeother \do\$\do\&\do\#\do\^\do\_\do\%\do\~}
\def\mn@doi{\begingroup\mn@urlcharsother \@ifnextchar [ {\mn@doi@}
  {\mn@doi@[]}}
\def\mn@doi@[#1]#2{\def\@tempa{#1}\ifx\@tempa\@empty \href
  {http://dx.doi.org/#2} {doi:#2}\else \href {http://dx.doi.org/#2} {#1}\fi
  \endgroup}
\def\mn@eprint#1#2{\mn@eprint@#1:#2::\@nil}
\def\mn@eprint@arXiv#1{\href {http://arxiv.org/abs/#1} {{\tt arXiv:#1}}}
\def\mn@eprint@dblp#1{\href {http://dblp.uni-trier.de/rec/bibtex/#1.xml}
  {dblp:#1}}
\def\mn@eprint@#1:#2:#3:#4\@nil{\def\@tempa {#1}\def\@tempb {#2}\def\@tempc
  {#3}\ifx \@tempc \@empty \let \@tempc \@tempb \let \@tempb \@tempa \fi \ifx
  \@tempb \@empty \def\@tempb {arXiv}\fi \@ifundefined
  {mn@eprint@\@tempb}{\@tempb:\@tempc}{\expandafter \expandafter \csname
  mn@eprint@\@tempb\endcsname \expandafter{\@tempc}}}

\bibitem[\protect\citeauthoryear{{Belloni}}{{Belloni}}{2010}]{2010LNP...794...53B}
{Belloni} T.~M.,  2010, {States and Transitions in Black Hole Binaries}.
Springer-Verlag Berlin Heidelberg, p.~53, \mn@doi{10.1007/978-3-540-76937-8\_3}

\bibitem[\protect\citeauthoryear{{Belloni} \& {Hasinger}}{{Belloni} \&
  {Hasinger}}{1990}]{1990A&A...227L..33B}
{Belloni} T.,  {Hasinger} G.,  1990, \aap, \href
  {https://ui.adsabs.harvard.edu/abs/1990A&A...227L..33B} {227, L33}

\bibitem[\protect\citeauthoryear{{Belloni} \& {Stella}}{{Belloni} \&
  {Stella}}{2014}]{2014SSRv..183...43B}
{Belloni} T.~M.,  {Stella} L.,  2014, \mn@doi [\ssr]
  {10.1007/s11214-014-0076-0}, \href
  {http://adsabs.harvard.edu/abs/2014SSRv..183...43B} {183, 43}

\bibitem[\protect\citeauthoryear{{Belloni}, {Motta}  \&
  {Mu{\~n}oz-Darias}}{{Belloni} et~al.}{2011}]{2011BASI...39..409B}
{Belloni} T.~M.,  {Motta} S.~E.,   {Mu{\~n}oz-Darias} T.,  2011, Bulletin of
  the Astronomical Society of India, \href
  {https://ui.adsabs.harvard.edu/abs/2011BASI...39..409B} {39, 409}

\bibitem[\protect\citeauthoryear{{Belloni}, {Zhang}, {Kylafis}, {Reig}  \&
  {Altamirano}}{{Belloni} et~al.}{2020}]{2020MNRAS.496.4366B}
{Belloni} T.~M.,  {Zhang} L.,  {Kylafis} N.~D.,  {Reig} P.,   {Altamirano} D.,
  2020, \mn@doi [\mnras] {10.1093/mnras/staa1843}, \href
  {https://ui.adsabs.harvard.edu/abs/2020MNRAS.496.4366B} {496, 4366}

\bibitem[\protect\citeauthoryear{{Bhuvana}, {Radhika}  \& {Nandi}}{{Bhuvana}
  et~al.}{2021}]{2021ATel14455....1B}
{Bhuvana} G.~R.,  {Radhika} D.,   {Nandi} A.,  2021, The Astronomer's Telegram,
  \href {https://ui.adsabs.harvard.edu/abs/2021ATel14455....1B} {14455, 1}

\bibitem[\protect\citeauthoryear{{Casella}, {Belloni}  \& {Stella}}{{Casella}
  et~al.}{2005}]{2005ApJ...629..403C}
{Casella} P.,  {Belloni} T.,   {Stella} L.,  2005, \mn@doi [\apj]
  {10.1086/431174}, \href {http://adsabs.harvard.edu/abs/2005ApJ...629..403C}
  {629, 403}

\bibitem[\protect\citeauthoryear{{Fender}, {Homan}  \& {Belloni}}{{Fender}
  et~al.}{2009}]{2009MNRAS.396.1370F}
{Fender} R.~P.,  {Homan} J.,   {Belloni} T.~M.,  2009, \mn@doi [\mnras]
  {10.1111/j.1365-2966.2009.14841.x}, \href
  {https://ui.adsabs.harvard.edu/abs/2009MNRAS.396.1370F} {396, 1370}

\bibitem[\protect\citeauthoryear{{Ferrigno}, {Grinberg}, {Bozzo}, {Savchenko},
  {Ducci}, {Wilms}, {Thalhammer}  \& {Rodriguez}}{{Ferrigno}
  et~al.}{2021}]{2021ATel14440....1F}
{Ferrigno} C.,  {Grinberg} V.,  {Bozzo} E.,  {Savchenko} V.,  {Ducci} L.,
  {Wilms} J.,  {Thalhammer} P.,   {Rodriguez} J.,  2021, The Astronomer's
  Telegram, \href {https://ui.adsabs.harvard.edu/abs/2021ATel14440....1F}
  {14440, 1}

\bibitem[\protect\citeauthoryear{{F{\"u}rst} et~al.,}{{F{\"u}rst}
  et~al.}{2015}]{2015ApJ...808..122F}
{F{\"u}rst} F.,  et~al., 2015, \mn@doi [\apj] {10.1088/0004-637X/808/2/122},
  \href {http://adsabs.harvard.edu/abs/2015ApJ...808..122F} {808, 122}

\bibitem[\protect\citeauthoryear{{F{\"u}rst} et~al.,}{{F{\"u}rst}
  et~al.}{2016}]{2016ApJ...828...34F}
{F{\"u}rst} F.,  et~al., 2016, \mn@doi [\apj] {10.3847/0004-637X/828/1/34},
  \href {https://ui.adsabs.harvard.edu/abs/2016ApJ...828...34F} {828, 34}

\bibitem[\protect\citeauthoryear{{Gao} et~al.,}{{Gao}
  et~al.}{2014}]{2014MNRAS.438..341G}
{Gao} H.~Q.,  et~al., 2014, \mn@doi [\mnras] {10.1093/mnras/stt2197}, \href
  {https://ui.adsabs.harvard.edu/abs/2014MNRAS.438..341G} {438, 341}

\bibitem[\protect\citeauthoryear{{Garc{\'\i}a}, {M{\'e}ndez}, {Karpouzas},
  {Belloni}, {Zhang}  \& {Altamirano}}{{Garc{\'\i}a}
  et~al.}{2021a}]{2021MNRAS.501.3173G}
{Garc{\'\i}a} F.,  {M{\'e}ndez} M.,  {Karpouzas} K.,  {Belloni} T.,  {Zhang}
  L.,   {Altamirano} D.,  2021a, \mn@doi [\mnras] {10.1093/mnras/staa3944},
  \href {https://ui.adsabs.harvard.edu/abs/2021MNRAS.501.3173G} {501, 3173}

\bibitem[\protect\citeauthoryear{{Garcia}, {Tomsick}, {Harrison}, {Connors}  \&
  {Mastroserio}}{{Garcia} et~al.}{2021b}]{2021ATel14352....1G}
{Garcia} J.,  {Tomsick} J.,  {Harrison} F.,  {Connors} R.,   {Mastroserio} G.,
  2021b, The Astronomer's Telegram, \href
  {https://ui.adsabs.harvard.edu/abs/2021ATel14352....1G} {14352, 1}

\bibitem[\protect\citeauthoryear{{Gendreau}, {Arzoumanian}  \&
  {Okajima}}{{Gendreau} et~al.}{2012}]{2012SPIE.8443E..13G}
{Gendreau} K.~C.,  {Arzoumanian} Z.,   {Okajima} T.,  2012, in Space Telescopes
  and Instrumentation 2012: Ultraviolet to Gamma Ray. p. 844313,
  \mn@doi{10.1117/12.926396}

\bibitem[\protect\citeauthoryear{{Husain}, {Mudambi}, {Garg}, {Misra}, {Sen}
  \& {Maqbool}}{{Husain} et~al.}{2021}]{2021ATel14400....1H}
{Husain} N.,  {Mudambi} S.~P.,  {Garg} A.,  {Misra} R.,  {Sen} S.,   {Maqbool}
  B.,  2021, The Astronomer's Telegram, \href
  {https://ui.adsabs.harvard.edu/abs/2021ATel14400....1H} {14400, 1}

\bibitem[\protect\citeauthoryear{{Ingram} \& {Motta}}{{Ingram} \&
  {Motta}}{2019}]{2019NewAR..8501524I}
{Ingram} A.~R.,  {Motta} S.~E.,  2019, \mn@doi [\nar]
  {10.1016/j.newar.2020.101524}, \href
  {https://ui.adsabs.harvard.edu/abs/2019NewAR..8501524I} {85, 101524}

\bibitem[\protect\citeauthoryear{{Ingram}, {Done}  \& {Fragile}}{{Ingram}
  et~al.}{2009}]{2009MNRAS.397L.101I}
{Ingram} A.,  {Done} C.,   {Fragile} P.~C.,  2009, \mn@doi [\mnras]
  {10.1111/j.1745-3933.2009.00693.x}, \href
  {https://ui.adsabs.harvard.edu/abs/2009MNRAS.397L.101I} {397, L101}

\bibitem[\protect\citeauthoryear{{Karpouzas}, {M{\'e}ndez}, {Ribeiro},
  {Altamirano}, {Blaes}  \& {Garc{\'\i}a}}{{Karpouzas}
  et~al.}{2020}]{2020MNRAS.492.1399K}
{Karpouzas} K.,  {M{\'e}ndez} M.,  {Ribeiro} E.~M.,  {Altamirano} D.,  {Blaes}
  O.,   {Garc{\'\i}a} F.,  2020, \mn@doi [\mnras] {10.1093/mnras/stz3502},
  \href {https://ui.adsabs.harvard.edu/abs/2020MNRAS.492.1399K} {492, 1399}

\bibitem[\protect\citeauthoryear{{Karpouzas}, {M{\'e}ndez}, {Garc{\'\i}a},
  {Zhang}, {Altamirano}, {Belloni}  \& {Zhang}}{{Karpouzas}
  et~al.}{2021}]{2021MNRAS.503.5522K}
{Karpouzas} K.,  {M{\'e}ndez} M.,  {Garc{\'\i}a} F.,  {Zhang} L.,  {Altamirano}
  D.,  {Belloni} T.,   {Zhang} Y.,  2021, \mn@doi [\mnras]
  {10.1093/mnras/stab827}, \href
  {https://ui.adsabs.harvard.edu/abs/2021MNRAS.503.5522K} {503, 5522}

\bibitem[\protect\citeauthoryear{{Kong}, {Charles}, {Kuulkers}  \&
  {Kitamoto}}{{Kong} et~al.}{2002}]{2002MNRAS.329..588K}
{Kong} A.~K.~H.,  {Charles} P.~A.,  {Kuulkers} E.,   {Kitamoto} S.,  2002,
  \mn@doi [\mnras] {10.1046/j.1365-8711.2002.05022.x}, \href
  {http://adsabs.harvard.edu/abs/2002MNRAS.329..588K} {329, 588}

\bibitem[\protect\citeauthoryear{{Kubota}, {Done}, {Tsurumi}  \&
  {Mizukawa}}{{Kubota} et~al.}{2023}]{2023arXiv230208314K}
{Kubota} A.,  {Done} C.,  {Tsurumi} K.,   {Mizukawa} R.,  2023, \mn@doi [arXiv
  e-prints] {10.48550/arXiv.2302.08314}, \href
  {https://ui.adsabs.harvard.edu/abs/2023arXiv230208314K} {p. arXiv:2302.08314}

\bibitem[\protect\citeauthoryear{{Kylafis}, {Reig}  \& {Papadakis}}{{Kylafis}
  et~al.}{2020}]{2020A&A...640L..16K}
{Kylafis} N.~D.,  {Reig} P.,   {Papadakis} I.,  2020, \mn@doi [\aap]
  {10.1051/0004-6361/202038468}, \href
  {https://ui.adsabs.harvard.edu/abs/2020A&A...640L..16K} {640, L16}

\bibitem[\protect\citeauthoryear{{Lamb} \& {Miller}}{{Lamb} \&
  {Miller}}{2001}]{2001ApJ...554.1210L}
{Lamb} F.~K.,  {Miller} M.~C.,  2001, \mn@doi [\apj] {10.1086/323148}, \href
  {https://ui.adsabs.harvard.edu/abs/2001ApJ...554.1210L} {554, 1210}

\bibitem[\protect\citeauthoryear{{Liu}, {Zhang}, {Tao}, {Chen}, {Qu}, {Zhang}
  \& {Zhang}}{{Liu} et~al.}{2021}]{2021ATel14504....1L}
{Liu} H.,  {Zhang} L.,  {Tao} L.,  {Chen} Y.,  {Qu} J.,  {Zhang} S.,   {Zhang}
  S.-n.,  2021, The Astronomer's Telegram, \href
  {https://ui.adsabs.harvard.edu/abs/2021ATel14504....1L} {14504, 1}

\bibitem[\protect\citeauthoryear{{Liu} et~al.,}{{Liu}
  et~al.}{2022}]{2022arXiv221109543L}
{Liu} H.,  et~al., 2022, arXiv e-prints, \href
  {https://ui.adsabs.harvard.edu/abs/2022arXiv221109543L} {p. arXiv:2211.09543}

\bibitem[\protect\citeauthoryear{{Mastichiadis}, {Petropoulou}  \&
  {Kylafis}}{{Mastichiadis} et~al.}{2022}]{2022A&A...662A.118M}
{Mastichiadis} A.,  {Petropoulou} M.,   {Kylafis} N.~D.,  2022, \mn@doi [\aap]
  {10.1051/0004-6361/202243397}, \href
  {https://ui.adsabs.harvard.edu/abs/2022A&A...662A.118M} {662, A118}

\bibitem[\protect\citeauthoryear{{Mastroserio}, {Garcia}, {Tomsick}, {Harrison}
   \& {Connors}}{{Mastroserio} et~al.}{2021}]{2021ATel14484....1M}
{Mastroserio} G.,  {Garcia} J.,  {Tomsick} J.,  {Harrison} F.,   {Connors} R.,
  2021, The Astronomer's Telegram, \href
  {https://ui.adsabs.harvard.edu/abs/2021ATel14484....1M} {14484, 1}

\bibitem[\protect\citeauthoryear{{McClintock} \& {Remillard}}{{McClintock} \&
  {Remillard}}{2006}]{2006csxs.book..157M}
{McClintock} J.~E.,  {Remillard} R.~A.,  2006, {Black hole binaries}.
Cambridge University Press, pp 157--213

\bibitem[\protect\citeauthoryear{{Miller-Jones} et~al.,}{{Miller-Jones}
  et~al.}{2012}]{2012MNRAS.421..468M}
{Miller-Jones} J.~C.~A.,  et~al., 2012, \mn@doi [\mnras]
  {10.1111/j.1365-2966.2011.20326.x}, \href
  {https://ui.adsabs.harvard.edu/abs/2012MNRAS.421..468M} {421, 468}

\bibitem[\protect\citeauthoryear{{Motta}, {Mu{\~n}oz-Darias}, {Casella},
  {Belloni}  \& {Homan}}{{Motta} et~al.}{2011}]{2011MNRAS.418.2292M}
{Motta} S.,  {Mu{\~n}oz-Darias} T.,  {Casella} P.,  {Belloni} T.,   {Homan} J.,
   2011, \mn@doi [\mnras] {10.1111/j.1365-2966.2011.19566.x}, \href
  {https://ui.adsabs.harvard.edu/abs/2011MNRAS.418.2292M} {418, 2292}

\bibitem[\protect\citeauthoryear{{Motta}, {Homan}, {Mu{\~n}oz Darias},
  {Casella}, {Belloni}, {Hiemstra}  \& {M{\'e}ndez}}{{Motta}
  et~al.}{2012}]{2012MNRAS.427..595M}
{Motta} S.,  {Homan} J.,  {Mu{\~n}oz Darias} T.,  {Casella} P.,  {Belloni}
  T.~M.,  {Hiemstra} B.,   {M{\'e}ndez} M.,  2012, \mn@doi [\mnras]
  {10.1111/j.1365-2966.2012.22037.x}, \href
  {https://ui.adsabs.harvard.edu/abs/2012MNRAS.427..595M} {427, 595}

\bibitem[\protect\citeauthoryear{{Nespoli}, {Belloni}, {Homan}, {Miller},
  {Lewin}, {M{\'e}ndez}  \& {van der Klis}}{{Nespoli}
  et~al.}{2003}]{2003A&A...412..235N}
{Nespoli} E.,  {Belloni} T.,  {Homan} J.,  {Miller} J.~M.,  {Lewin} W.~H.~G.,
  {M{\'e}ndez} M.,   {van der Klis} M.,  2003, \mn@doi [\aap]
  {10.1051/0004-6361:20031423}, \href
  {https://ui.adsabs.harvard.edu/abs/2003A&A...412..235N} {412, 235}

\bibitem[\protect\citeauthoryear{{Nobili}, {Turolla}, {Zampieri}  \&
  {Belloni}}{{Nobili} et~al.}{2000}]{2000ApJ...538L.137N}
{Nobili} L.,  {Turolla} R.,  {Zampieri} L.,   {Belloni} T.,  2000, \mn@doi
  [\apjl] {10.1086/312810}, \href
  {https://ui.adsabs.harvard.edu/abs/2000ApJ...538L.137N} {538, L137}

\bibitem[\protect\citeauthoryear{{Pal}, {Mandal}, {Hazra}, {Saha}  \&
  {Ghanta}}{{Pal} et~al.}{2021}]{2021ATel14351....1P}
{Pal} S.,  {Mandal} M.,  {Hazra} M.,  {Saha} D.,   {Ghanta} A.,  2021, The
  Astronomer's Telegram, \href
  {https://ui.adsabs.harvard.edu/abs/2021ATel14351....1P} {14351, 1}

\bibitem[\protect\citeauthoryear{{Peirano}, {M{\'e}ndez}, {Garc{\'\i}a}  \&
  {Belloni}}{{Peirano} et~al.}{2023}]{2023MNRAS.519.1336P}
{Peirano} V.,  {M{\'e}ndez} M.,  {Garc{\'\i}a} F.,   {Belloni} T.,  2023,
  \mn@doi [\mnras] {10.1093/mnras/stac3553}, \href
  {https://ui.adsabs.harvard.edu/abs/2023MNRAS.519.1336P} {519, 1336}

\bibitem[\protect\citeauthoryear{{Sguera} et~al.,}{{Sguera}
  et~al.}{2021}]{2021ATel14354....1S}
{Sguera} V.,  et~al., 2021, The Astronomer's Telegram, \href
  {https://ui.adsabs.harvard.edu/abs/2021ATel14354....1S} {14354, 1}

\bibitem[\protect\citeauthoryear{{Stella} \& {Vietri}}{{Stella} \&
  {Vietri}}{1998}]{1998ApJ...492L..59S}
{Stella} L.,  {Vietri} M.,  1998, \mn@doi [\apjl] {10.1086/311075}, \href
  {http://adsabs.harvard.edu/abs/1998ApJ...492L..59S} {492, L59}

\bibitem[\protect\citeauthoryear{{Stella} \& {Vietri}}{{Stella} \&
  {Vietri}}{1999}]{1999PhRvL..82...17S}
{Stella} L.,  {Vietri} M.,  1999, \mn@doi [\prl] {10.1103/PhysRevLett.82.17},
  \href {https://ui.adsabs.harvard.edu/abs/1999PhRvL..82...17S} {82, 17}

\bibitem[\protect\citeauthoryear{{Stiele} \& {Kong}}{{Stiele} \&
  {Kong}}{2017}]{2017ApJ...844....8S}
{Stiele} H.,  {Kong} A.~K.~H.,  2017, \mn@doi [\apj]
  {10.3847/1538-4357/aa774e}, \href
  {http://adsabs.harvard.edu/abs/2017ApJ...844....8S} {844, 8}

\bibitem[\protect\citeauthoryear{{Tagger} \& {Pellat}}{{Tagger} \&
  {Pellat}}{1999}]{1999A&A...349.1003T}
{Tagger} M.,  {Pellat} R.,  1999, \aap, \href
  {https://ui.adsabs.harvard.edu/abs/1999A&A...349.1003T} {349, 1003}

\bibitem[\protect\citeauthoryear{{Titarchuk} \& {Osherovich}}{{Titarchuk} \&
  {Osherovich}}{1999}]{1999ApJ...518L..95T}
{Titarchuk} L.,  {Osherovich} V.,  1999, \mn@doi [\apjl] {10.1086/312083},
  \href {https://ui.adsabs.harvard.edu/abs/1999ApJ...518L..95T} {518, L95}

\bibitem[\protect\citeauthoryear{{Wang} et~al.,}{{Wang}
  et~al.}{2021a}]{2021ATel14384....1W}
{Wang} J.,  et~al., 2021a, The Astronomer's Telegram, \href
  {https://ui.adsabs.harvard.edu/abs/2021ATel14384....1W} {14384, 1}

\bibitem[\protect\citeauthoryear{{Wang} et~al.,}{{Wang}
  et~al.}{2021b}]{2021ATel14490....1W}
{Wang} J.,  et~al., 2021b, The Astronomer's Telegram, \href
  {https://ui.adsabs.harvard.edu/abs/2021ATel14490....1W} {14490, 1}

\bibitem[\protect\citeauthoryear{{Wijnands}, {Homan}  \& {van der
  Klis}}{{Wijnands} et~al.}{1999}]{1999ApJ...526L..33W}
{Wijnands} R.,  {Homan} J.,   {van der Klis} M.,  1999, \mn@doi [\apjl]
  {10.1086/312365}, \href
  {https://ui.adsabs.harvard.edu/abs/1999ApJ...526L..33W} {526, L33}

\bibitem[\protect\citeauthoryear{{Zhang}, {Wang}, {M{\'e}ndez}, {Chen}, {Qu},
  {Altamirano}  \& {Belloni}}{{Zhang} et~al.}{2017}]{2017ApJ...845..143Z}
{Zhang} L.,  {Wang} Y.,  {M{\'e}ndez} M.,  {Chen} L.,  {Qu} J.,  {Altamirano}
  D.,   {Belloni} T.,  2017, \mn@doi [\apj] {10.3847/1538-4357/aa8138}, \href
  {https://ui.adsabs.harvard.edu/abs/2017ApJ...845..143Z} {845, 143}

\makeatother
\end{thebibliography}




\appendix

\section{Parameters of power density spectra for each snapshot}

\begin{table*}
\caption{Parameters of the BLN components of the PDS}
\begin{center}
\scriptsize
\begin{tabular}{llrrrrrrrrr}
\hline\noalign{\smallskip}
 \multicolumn{1}{c}{Snapshot} & \multicolumn{1}{c}{Obs.} & \multicolumn{1}{c}{$\nu_{\mr{BLN1}}$ [Hz]} & \multicolumn{1}{c}{rms$_{\mr{BLN1}}$ } & \multicolumn{1}{c}{$\nu_{\mr{BLN2}}$ [Hz]} & \multicolumn{1}{c}{rms$_{\mr{BLN2}}$ } & \multicolumn{1}{c}{$\nu_{\mr{0;PN}}$ [Hz]}  & \multicolumn{1}{c}{$\Delta_{\mr{PN}}$ [Hz] }  & \multicolumn{1}{c}{rms$_{\mr{PN}}$ } &  \multicolumn{1}{c}{Q$_{\mr{PN}}$}  & \multicolumn{1}{c}{$\sigma_{\mr{PN}}$} \\
\hline\noalign{\smallskip}
1	&4& $2.1452_{-0.1819}^{+0.1835} $& $	0.0812_{-0.0021}^{+0.0019} $& $				$& $				    $& $0.3735_{-0.0211}^{+0.0213} $& $	0.2414_{-0.0369}^{+0.0459} $& $	0.0440_{-0.0032}^{+0.0033} $& $	0.77$& $	6.88	$\\
\smallskip
2	&4&$ 2.1322_{-0.1435}^{+0.1760} $& $	0.0821_{-0.0023}^{+0.002} $& $				$& $				    $& $0.3813_{-0.0192}^{+0.0203} $& $	0.2146_{-0.0312}^{+0.0373} $& $	0.0452_{-0.0028}^{+0.0031} $& $	0.89$& $	8.07	$\\
\smallskip
3	&4&$ 2.3240_{-0.1943}^{+0.1669} $& $	0.0826_{-0.0014}^{+0.0013} $& $				$& $				    $& $0.6567_{-0.0111}^{+0.0137} $& $	0.0898_{-0.0220}^{+0.0227} $& $	0.0322_{-0.0035}^{+0.0031} $& $	3.66$& $	4.60	$\\
\smallskip
4	&4&$ 2.5522_{-0.1145}^{+0.1596} $& $	0.0951_{-0.0018}^{+0.0012} $& $				$& $				    $& $0.5847_{-0.0152}^{+0.0171} $& $	0.1148_{-0.0216}^{+0.0294} $& $	0.0358_{-0.0035}^{+0.0048} $& $	2.55$& $	5.11	$\\
\smallskip
5	&4&$ 2.4371_{-0.1798}^{+0.2075} $& $	0.0842_{-0.0020}^{+0.0019} $& $				$& $				    $& $0.5839_{-0.0511}^{+0.0428} $& $	0.1789_{-0.0560}^{+0.0543} $& $	0.0315_{-0.0051}^{+0.0047} $& $	1.63$& $	3.09	$\\
\smallskip
6	&4&$ 2.2953_{-0.2241}^{+0.2518} $& $	0.0733_{-0.0033}^{+0.0032} $& $				$& $				    $& $0.5459_{-0.0244}^{+0.0266} $& $	0.3375_{-0.0608}^{+0.0608} $& $	0.0494_{-0.0047}^{+0.0042} $& $	0.81$& $	5.26	$\\
\smallskip
7	&4&$ 1.8336_{-0.1689}^{+0.2090} $& $	0.0751_{-0.0026}^{+0.0022} $& $				$& $				    $& $0.4518_{-0.0260}^{+0.0265} $& $	0.1897_{-0.0443}^{+0.0614} $& $	0.0340_{-0.0041}^{+0.0042} $& $	1.19$& $	4.15	$\\
\smallskip
8	&4&$ 2.7912_{-0.3074}^{+0.3422} $& $	0.0697_{-0.0040}^{+0.0039} $& $				$& $				    $& $0.5726_{-0.0554}^{+0.0584} $& $	0.4279_{-0.1036}^{+0.0721} $& $	0.0382_{-0.0067}^{+0.0052} $& $	0.67$& $	2.85	$\\
\smallskip
9	&4&$ 3.1838_{-0.3497}^{+0.3432} $& $	0.0680_{-0.0027}^{+0.0030} $& $				$& $				    $& $0.5792_{-0.0440}^{+0.0380} $& $	0.3171_{-0.0819}^{+0.0820} $& $	0.0365_{-0.0046}^{+0.0032} $& $	0.91$& $	3.97	$\\
\smallskip
10	&4&$ 3.3254_{-0.6869}^{+0.6909} $& $	0.0400_{-0.0047}^{+0.0038} $& $				$& $				    $& $0.5827_{-0.0341}^{+0.0298} $& $	0.4421_{-0.0576}^{+0.0506} $& $	0.0428_{-0.0030}^{+0.0025} $& $	0.66$& $	7.13	$\\
\smallskip
11	&4&$ 2.5000_{-0.0337}^{+0.0309} $& $		0.0832\pm0.0005 $& $				$& $				    $& $			$& $				 $& $				$& $	$& 		\\
\smallskip
12	&4&$ 2.7307_{-0.2869}^{+0.2485} $& $	0.0710_{-0.0030}^{+0.0024} $& $				$& $				    $& $0.6734_{-0.0539}^{+0.0340} $& $	0.2518_{-0.0917}^{+0.1400} $& $	0.0263_{-0.0039}^{+0.0042} $& $	1.34$& $	3.37	$\\
\smallskip
13	&4&$ 2.0262_{-0.1048}^{+0.1219} $& $	0.0669_{-0.0010}^{+0.0011} $& $				$& $				    $& $			$& $				$& $				$& $	$& 			\\
\smallskip
14	&4&$ 1.3090_{-0.1240}^{+0.1311} $& $	0.0524_{-0.0020}^{+0.0019} $& $				$& $				    $& $			$& $				$& $				$& $	$& 			\\
\smallskip
15	&5&$ 2.0091_{-0.1302}^{+0.1242} $& $	0.0581_{-0.0016}^{+0.0013} $& $				$& $				    $& $1.1683_{-0.0173}^{+0.0237} $& $	0.0605_{-0.0300}^{+0.0330} $& $	0.0133_{-0.0019}^{+0.0020} $& $	9.66$& $	3.50$	\\
\smallskip
16	&5&$ 1.8510_{-0.0617}^{+0.0749} $& $	0.0479\pm0.0002 $& $				$& $				    $& $1.2605_{-0.0017}^{+0.1133} $& $	0.0004_{-0.0004}^{+0.0332} $& $	0.0111_{-0.0025}^{+0.0027} $& $	1575.63$& $	2.22$	\\
\smallskip
17	&5&$ 2.3319_{-0.1402}^{+0.1437} $& $	0.0508\pm0.0011 $& $				$& $				    $& $			$& $				$& $				$& $	$& 			\\
\smallskip
18	&5&$ 2.1353_{-0.1173}^{+0.0829} $& $	0.0477_{-0.0010}^{+0.0004} $& $				$& $				    $& $1.3287_{-0.0680}^{+0.0408} $& $	0.1428_{-0.0665}^{+0.1790} $& $	0.0109_{-0.0015}^{+0.0020} $& $	4.65$& $	3.63$	\\
\smallskip
19	&5&$ 2.1492_{-0.0898}^{+0.0944} $& $	0.0481\pm0.0007 $& $				$& $				    $& $			$& $				$& $				$& $	$&	\\\smallskip
20	&5&$ 2.0416_{-0.0954}^{+0.0991} $& $	0.0513\pm0.0008 $& $				$& $				    $& $			$& $				$& $				$& $	$&			\\
\smallskip
21	&5&$ 2.1487_{-0.0975}^{+0.0996}$& $	0.0530_{-0.0008}^{+0.0007} $& $				$& $				    $& $			$& $				$& $				$& $	$&			\\
\smallskip
22	&5&$ 2.1986_{-0.1387}^{+0.1171} $& $	0.0432_{-0.0015}^{+0.0009} $& $				$& $				    $& $1.3917_{-0.1363}^{+0.0846} $& $	0.2784_{-0.1122}^{+0.2228} $& $	0.0122_{-0.0023}^{+0.0032} $& $	2.50$& $	2.65$	\\
\smallskip
23	&5&$ 0.0767_{-0.0259}^{+0.6209} $& $	0.0157_{-0.0056}^{+0.0027} $& $	0.0062_{-0.0062}^{+0.0090} $& $	0.0189_{-0.0034}^{+0.1264} $& $1.6945_{-0.0970}^{+0.0852} $& $	0.3321_{-0.1637}^{+0.1679} $& $	0.0125_{-0.0039}^{+0.0020} $& $	2.55$& $	1.60$	\\
\smallskip
24	&5&$ 2.1943_{-0.1216}^{+0.1317} $& $	0.0411_{-0.0012}^{+0.0011} $& $				$& $				    $& $1.7526_{-0.0605}^{+0.0632} $& $	0.3513_{-0.0882}^{+0.1049} $& $	0.0163_{-0.0020}^{+0.0020} $& $	2.49$& $	4.08$	\\
\smallskip
25	&5&$ 2.5487_{-0.5816}^{+0.8517} $& $	0.0162_{-0.0014}^{+0.0018} $& $	0.0152_{-0.0033}^{+0.0040} $& $	0.0114_{-0.0011}^{+0.0010} $& $				$& $				$& $				$& $	$& 		\\
\smallskip
26	&5&$ 1.9484_{-1.9484}^{+8.0516} $& $	0.0144_{-0.0144}^{+0.0042} $& $	0.0155_{-0.0022}^{+0.0029} $& $	0.0294_{-0.0023}^{+0.0021} $& $2.5183_{-0.5183}^{+0.1826} $& $0.6991_{-0.3114}^{+0.8009} $& $0.01540_{-0.0036}^{+0.0000} $& $1.80	$& $2.14$		\\	
\smallskip
27	&5&$ 0.0311_{-0.0039}^{+0.0046} $& $	0.0238_{-0.0015}^{+0.0014} $& $				$& $				    $& $2.3748_{-0.1761}^{+0.1280} $& $	2.1848_{-0.2014}^{+0.2126} $& $	0.0297\pm0.0008 $& $	0.54$& $	18.56$	\\
\smallskip
28	&6&$ 3.4628_{-1.2398}^{+1.5372} $& $	0.0139_{-0.0024}^{+0.0026} $& $	0.0241_{-0.0056}^{+0.0080} $& $	0.0102_{-0.0011}^{+0.0010} $& $				$& $				$& $				$& $	$& 			\\
\smallskip
29	&6&$ 2.4158_{-0.1240}^{+0.1343} $& $	0.0466_{-0.0008}^{+0.0009} $& $				$& $				    $& $			$& $				$& $				$& $	$& 			\\
\smallskip
30	&6&$ 3.1303_{-0.2570}^{+0.2757} $& $	0.0287\pm0.0010 $& $	0.0177_{-0.0024}^{+0.0030} $& $	0.0299_{-0.0019}^{+0.0018} $& $			$& $				$& $				$& $	$& 		\\	
\smallskip
31	&6&$ 0.4151_{-0.2069}^{+0.4900} $& $	0.0115_{-0.0025}^{+0.0019} $& $	0.0359_{-0.0065}^{+0.0077} $& $	0.0206_{-0.0018}^{+0.0014} $& $2.6181_{-0.2205}^{+0.2026} $& $	1.3562_{-0.1902}^{+0.1966} $& $	0.0254_{-0.0018}^{+0.0017} $& $	0.97$& $	7.06$	\\
\smallskip
32	&6&$ 2.4174_{-0.0865}^{+0.0930} $& $	0.0474\pm0.0006 $& $				$& $				    $& $			$& $				$& $				$& $	$& 			\\
\smallskip
33	&6&$ 2.4658_{-0.1159}^{+0.1302} $& $	0.0474\pm0.0007 $& $				$& $				    $& $1.3166_{-0.0841}^{+0.0524} $& $	0.1375_{-0.0594}^{+0.0948} $& $	0.0116_{-0.0021}^{+0.0019} $& $	4.79$& $	2.76$	\\
\smallskip
34	&6&$ 2.2042_{-0.1346}^{+0.0984} $& $	0.0512_{-0.0011}^{+0.0007} $& $				$& $				    $& $			$& $				$& $				$& $	$& 			\\
\smallskip
35	&6&$ 2.0021_{-0.0951}^{+0.0968} $& $	0.048_{-0.0008}^{+0.0007} $& $				$& $				    $& $			$& $				$& $					$& 			\\
\smallskip
36	&6&$ 0.4962_{-0.1370}^{+0.2456} $& $	0.0145_{-0.0016}^{+0.0022} $& $	0.0035_{-0.0035}^{+0.0040} $& $	0.0175_{-0.0045}^{+0.0508}  $& $2.3139_{-0.0950}^{+0.1033} $& $	1.2395_{-0.1232}^{+0.1120} $& $	0.0304_{-0.0014}^{+0.0010} $& $	0.93$& $	10.86$	\\
\smallskip
37	&6&$ 4.2463_{-0.5145}^{+0.5275} $& $	0.0238_{-0.0015}^{+0.0013} $& $	0.0264_{-0.0026}^{+0.0030} $& $	0.0275_{-0.0014}^{+0.0013}  $& $			$& $				$& $				$& $	$& 			\\
\smallskip
38	&7&$ 5.1465_{-1.4294}^{+1.5864} $& $	0.0160_{-0.0024}^{+0.0021} $& $	0.0308_{-0.0042}^{+0.0051} $& $	0.0169\pm0.0010    $& $			$& $				$& $				$& $	$& 			\\
\smallskip
39	&7&$ 0.3791_{-0.1371}^{+0.2590} $& $	0.0086_{-0.0019}^{+0.0018} $& $	0.0169_{-0.0028}^{+0.0019} $& $	0.0278_{-0.0010}^{+0.0009}   $& $			$& $				$& $				$& $	$& 			\\
\smallskip
40	&7&$ 1.7587_{-0.4063}^{+0.3483} $& $	0.0095_{-0.0014}^{+0.0012} $& $	0.0131_{-0.0027}^{+0.0023} $& $	0.0208_{-0.0011}^{+0.0012}  $& $			$& $				$& $				$& $	$& 			\\
\smallskip
41	&7&$ 6.7280_{-0.2861}^{+0.2661} $& $	0.0408_{-0.0008}^{+0.0007} $& $	0.0573_{-0.0072}^{+0.0091} $& $	0.0161\pm0.0006  $& $			$& $				$& $				$& $	$& 	             	\\
\smallskip
42	&7&$ 0.1150_{-0.0290}^{+0.0333} $& $	0.0084_{-0.0013}^{+0.0011} $& $				$& $				    $& $1.2728_{-0.1035}^{+0.0926} $& $	1.6399_{-0.1061}^{+0.1170} $& $	0.0359_{-0.0000}^{+6.0677} $& $	0.39$& 	\\
\smallskip
43	&7&$ 1.3713_{-0.7442}^{+0.6516} $& $	0.0193_{-0.0084}^{+0.0072} $& $	0.0976_{-0.0446}^{+0.0912} $& $	0.0073_{-0.0015}^{+0.0022} $& $2.0526_{-0.2463}^{+0.2251} $& $	1.4155_{-0.2951}^{+0.2026} $& $	0.0305_{-0.0055}^{+0.0054} $& $	0.73$& $	2.77$	\\
\smallskip
44	&7&$ 0.0281_{-0.0032}^{+0.0035} $& $	0.0286\pm0.0015 $& $				$& $				    $& $2.0022_{-0.1488}^{+0.1369} $& $	1.4778_{-0.2232}^{+0.2659} $& $	0.0250\pm0.0013 $& $	0.68$& $	9.62$	\\
\smallskip
45	&7&$ 0.0292_{-0.0031}^{+0.0036} $& $	0.0229_{-0.0012}^{+0.0011} $& $				$& $				    $& $			$& $				$& $				$& $	$& 			\\
\smallskip
46a	&7&$ 0.0179_{-0.0053}^{+0.0070 } $& $	0.0088_{-0.0011}^{+0.0010 } $& $				$& $				    $& $			$& $				$& $					$& 			\\
\smallskip
46b	&7&$ 0.0853_{-0.0146}^{+0.0178} $& $	0.0146\pm0.0006 $& $				$& $				    $& $			$& $				$& $					$& 			\\
\smallskip
47a	&7&$ 0.0148_{-0.0053}^{+0.0051} $& $	0.0063_{-0.0010}^{+0.0008} $& $				$& $				    $& $			$& $				$& $					$& 			\\
\smallskip
47b	&7&$ 0.0154_{-0.0039}^{+0.0041} $& $	0.0091_{-0.0006}^{+0.0009} $& $				$& $				    $& $			$& $				$& $					$& 			\\
\smallskip
48	&7&$ 0.0972_{-0.0261}^{+0.0333} $& $	0.0074\pm0.0006 $& $				$& $				    $& $0.2872_{-0.0244}^{+0.0210} $& $	0.0413_{-0.0180}^{+0.0278} $& $	0.0047_{-0.0008}^{+0.0007} $& $	3.48$& $	2.94$	\\
\smallskip
49	&7&$ 0.0238_{-0.0053}^{+0.0067} $& $	0.0091\pm0.0007 $& $				$& $				    $& $			$& $				$& $					$& 			\\
\hline\noalign{\smallskip} 
\end{tabular} 
\end{center}
\label{Tab:PDS}
\end{table*}

\begin{table*}
\addtocounter{table}{-1}
\caption{continued}
\begin{center}
\scriptsize
\begin{tabular}{llrrrrrrrrr}
\hline\noalign{\smallskip}
 \multicolumn{1}{c}{Snapshot} & \multicolumn{1}{c}{Obs.} & \multicolumn{1}{c}{$\nu_{\mr{BLN1}}$ [Hz]} & \multicolumn{1}{c}{rms$_{\mr{BLN1}}$ } & \multicolumn{1}{c}{$\nu_{\mr{BLN2}}$ [Hz]} & \multicolumn{1}{c}{rms$_{\mr{BLN2}}$ } & \multicolumn{1}{c}{$\nu_{\mr{0;PN}}$ [Hz]}  & \multicolumn{1}{c}{$\Delta_{\mr{PN}}$ [Hz] }  & \multicolumn{1}{c}{rms$_{\mr{PN}}$ } &  \multicolumn{1}{c}{Q$_{\mr{PN}}$}  & \multicolumn{1}{c}{$\sigma_{\mr{PN}}$} \\
\hline\noalign{\smallskip}
50	&7&$ 0.0489_{-0.0099}^{+0.0125} $& $	0.0082_{-0.0007}^{+0.0006} $& $				$& $				    $& $			$& $				$& $					$& 			\\
\smallskip
51	&7&$ 0.0241_{-0.0045}^{+0.0055} $& $	0.0078\pm0.0006 $& $				$& $				    $& $			$& $				$& $					$& 			\\
\smallskip
52	&8&$ 0.0331_{-0.0096}^{+0.0258} $& $	0.0096_{-0.0009}^{+0.0008} $& $				$& $				    $& $			$& $				$& $					$& 			\\
\smallskip
53	&8&$ 0.0331_{-0.0096}^{+0.0258} $& $	0.0096_{-0.0009}^{+0.0008} $& $				$& $				    $& $			$& $				$& $					$& 			\\
\hline\noalign{\smallskip} 
\end{tabular} 
\end{center}
\label{Tab:PDS}
\end{table*}

\begin{table*}
\caption{Parameters of the fundamental QPOs}
\begin{center}
\footnotesize
\begin{tabular}{llrrrrr}
\hline\noalign{\smallskip}
 \multicolumn{1}{c}{Snapshot} & \multicolumn{1}{c}{Obs.} & \multicolumn{1}{c}{$\nu_{\mr{0;QPO}}$ [Hz]}  & \multicolumn{1}{c}{$\Delta_{\mr{QPO}}$ [Hz] }  & \multicolumn{1}{c}{rms$_{\mr{QPO}}$ } &  \multicolumn{1}{c}{Q$_{\mr{QPO}}$}  & \multicolumn{1}{c}{$\sigma_{\mr{QPO}}$} \\
\hline\noalign{\smallskip}
1	&4& $3.1927_{-0.0287}^{+0.0277} $& $	0.1305_{-0.0545}^{+0.0571} $& $	0.0309_{-0.0028}^{+0.0026} $& $12.23	$& $5.52	$\\
\smallskip
2	&4& $3.0586_{-0.0320}^{+0.0354} $& $	0.1117_{-0.0341}^{+0.0306} $& $	0.0314_{-0.0023}^{+0.0020} $& $13.69	$& $6.83	$\\
\smallskip
3	&4& $3.2054_{-0.0183}^{+0.0197} $& $	0.1083_{-0.0243}^{+0.0312} $& $	0.0297\pm0.0019 $& $14.80	$& $7.82	$\\
\smallskip
4	&4& $3.1645_{-0.0058}^{+0.0001} $& $	0.0004_{-0.0004}^{+0.0139} $& $	0.0262_{-0.0030}^{+0.0029} $& $113.83	$& $4.37	$\\
\smallskip
5	&4& $3.2709_{-0.0332}^{+0.0371} $& $	0.1388_{-0.0475}^{+0.0538} $& $	0.0318_{-0.0029}^{+0.0026} $& $11.78	$& $5.48	$\\
\smallskip
6	&4& $3.3433_{-0.0226}^{+0.0255} $& $	0.1424_{-0.0235}^{+0.0235} $& $	0.0313_{-0.0015}^{+0.0014} $& $11.74	$& $10.43	$\\
\smallskip
7	&4& $3.5808_{-0.0345}^{+0.0416} $& $	0.1945_{-0.0269}^{+0.0282} $& $	0.0296_{-0.0021}^{+0.0020} $& $9.21	$& $7.05	$\\
\smallskip
8	&4& $3.8551_{-0.0240}^{+0.1216} $& $	0.0985_{-0.0924}^{+0.1615} $& $	0.0230_{-0.0030}^{+0.0024} $& $19.57	$& $3.83	$\\
\smallskip
9	&4& $3.9390_{-0.0431}^{+0.0631} $& $	0.1751_{-0.0412}^{+0.0394} $& $	0.0265_{-0.0025}^{+0.0020} $& $11.25	$& $5.30	$\\
\smallskip
10	&4& $4.5179_{-0.0362}^{+0.0370} $& $	0.3839_{-0.0778}^{+0.1100} $& $	0.0270_{-0.0018}^{+0.0020} $& $5.88	$& $7.50	$\\
\smallskip
11	&4& $4.7028_{-0.0169}^{+0.0000} $& $    \le 0.0289                              $& $	0.0135_{-0.0019}^{+0.0017} $& $81.36	$& $3.55	$\\
\smallskip
12	&4& $4.2365_{-0.0512}^{+0.0279} $& $	0.1858_{-0.1318}^{+0.0985} $& $	0.0222_{-0.0027}^{+0.0023} $& $11.40	$& $4.11	$\\
\smallskip
13	&4& $4.7166_{-0.0378}^{+0.0275} $& $	0.2018_{-0.0738}^{+0.0748} $& $	0.0203_{-0.0015}^{+0.0013} $& $11.69	$& $6.77	$\\
\smallskip
14	&4& $5.3756_{-0.0940}^{+0.1007} $& $	0.5180_{-0.1134}^{+0.1443} $& $	0.0221_{-0.0019}^{+0.0018} $& $5.19	$& $5.82	$\\
\smallskip
15	&5& $5.1222_{-0.0865}^{+0.0603} $& $	0.4805_{-0.1822}^{+0.2265} $& $	0.0218_{-0.0029}^{+0.0034} $& $5.33	$& $3.76	$\\
\smallskip
16	&5& $5.1222			             $& $	0.4805		                	    $& $	0.0268_{-0.0014}^{+0.0013} $& $	5.33	$& $9.57	$\\
\smallskip
17	&5& $6.0055_{-0.1311}^{+0.1371} $& $	0.4047_{-0.1483}^{+0.2025} $& $	0.0147_{-0.0022}^{+0.0021} $& $7.42	$& $3.34	$\\
\smallskip
18	&5& $5.7306_{-0.3608}^{+0.2707} $& $	1.1050_{-0.3482}^{+0.4238} $& $	0.0143_{-0.0016}^{+0.0015} $& $2.59	$& $4.47	$\\
\smallskip
20	&5& $6.0474_{-0.0894}^{+0.0894} $& $	0.3689_{-0.1232}^{+0.1934} $& $	0.0145_{-0.0015}^{+0.0014} $& $8.20	$& $4.83	$\\
\smallskip
21	&5& $6.0163_{-0.1291}^{+0.1523} $& $	0.2076_{-0.1226}^{+0.1147} $& $	0.0120\pm0.0017 $& $14.49	$& $3.53	$\\
\smallskip
23	&5& $5.7170_{-0.1528}^{+0.0882} $& $	0.2367_{-0.2367}^{+0.2932} $& $	0.0101_{-0.0026}^{+0.0028} $& $12.08	$& $1.94	$\\
\smallskip
25	&5& $5.0702_{-0.1029}^{+0.0971} $& $	0.2769_{-0.1427}^{+0.0848} $& $	0.0164_{-0.0017}^{+0.0012} $& $9.16	$& $4.82	$\\
\smallskip
26	&5& $5.5746_{-0.1694}^{+0.1538} $& $	0.5527_{-0.1708}^{+0.1594} $& $	0.0165_{-0.0020}^{+0.0018} $& $5.04	$& $4.13	$\\
\smallskip
28	&6& $4.7980_{-0.0446}^{+0.0543} $& $	0.2381_{-0.1039}^{+0.0944} $& $	0.0168_{-0.0016}^{+0.0015} $& $10.08	$& $5.25	$\\
\smallskip
30	&6& $5.3433_{-0.1422}^{+0.1649} $& $	0.7102_{-0.1493}^{+0.1844} $& $	0.0127_{-0.0014}^{+0.0013} $& $3.76	$& $4.54	$\\
\smallskip
34	&6& $6.1004_{-0.2141}^{+0.1743} $& $	0.5000_{-0.1262}^{+0.2976} $& $	0.0119_{-0.0013}^{+0.0025} $& $6.10	$& $4.58	$\\
\smallskip
35	&6& $6.3872_{-0.1499}^{+0.1728} $& $	0.5049_{-0.2439}^{+0.3913} $& $	0.0113_{-0.0018}^{+0.0017} $& $6.33	$& $3.14	$\\
\smallskip
37	&6& $5.5273_{-0.0576}^{+0.0557} $& $	0.3409_{-0.0800}^{+0.1297} $& $	0.0144\pm0.0011 $& $8.11	$& $6.55	$\\
\smallskip
38	&7& $5.1369_{-0.0518}^{+0.0398} $& $	0.3465_{-0.0685}^{+0.0768} $& $	0.0183\pm0.0010 $& $7.41	$& $9.15	$\\
\smallskip
39	&7& $5.3578_{-0.1368}^{+0.1439} $& $	0.6596_{-0.1345}^{+0.1496} $& $	0.0182_{-0.0016}^{+0.0014} $& $4.06	$& $5.69	$\\
\smallskip
40	&7& $4.9980_{-0.0491}^{+0.0647} $& $	0.4106_{-0.0445}^{+0.0532} $& $	0.0177_{-0.0009}^{+0.0012} $& $6.09	$& $9.83	$\\
\smallskip
44	&7& $5.6652_{-0.1501}^{+0.1281} $& $	0.5509_{-0.2161}^{+0.2121} $& $	0.0137_{-0.0021}^{+0.0018} $& $5.14	$& $3.26	$\\
\smallskip
45	&7& $5.3499_{-0.1033}^{+0.1392} $& $	0.4247_{-0.1172}^{+0.1133} $& $	0.0156_{-0.0019}^{+0.0017} $& $6.30	$& $4.11	$\\
\smallskip
46a	&7& $5.0346_{-0.0471}^{+0.0472} $& $	0.3165_{-0.0558}^{+0.0788} $& $	0.0189_{-0.0011}^{+0.0010} $& $	7.95	$& $	8.59	$\\
\smallskip
46b	&7& $5.1499_{-0.0946}^{+0.1034} $& $	0.4475_{-0.1134}^{+0.1172} $& $	0.0185_{-0.0016}^{+0.0014} $& $	5.75	$& $	5.78	$\\
\smallskip
47a	&7& $4.9487_{-0.0588}^{+0.0564} $& $	0.2548_{-0.0411}^{+0.0549} $& $	0.0164\pm0.0011 $& $	9.71	$& $	7.45	$\\
\smallskip
47b	&7& $4.9740_{-0.0377}^{+0.0389} $& $	0.3144_{-0.0418}^{+0.0511} $& $	0.0195\pm0.0009 $& $	7.91	$& $	10.83$\\	
\smallskip
48	&7& $4.9100_{-0.0458}^{+0.0503} $& $	0.2545_{-0.0400}^{+0.0470} $& $    0.0164\pm0.0009 $& $9.65	$& $9.11	$\\
\smallskip
49	&7& $5.0274_{-0.0376}^{+0.0363} $& $	0.3204_{-0.0441}^{+0.0505} $& $	0.0170\pm0.0007 $& $7.85	$& $12.14	$\\
\smallskip
50	&7& $4.8690_{-0.0433}^{+0.0456} $& $	0.2569_{-0.0532}^{+0.0469} $& $	0.0170\pm0.0008 $& $9.48	$& $10.63	$\\
\smallskip
51	&7& $4.8514_{-0.0641}^{+0.0773} $& $	0.3071_{-0.0591}^{+0.0517} $& $	0.0162\pm0.0008 $& $7.90	$& $10.13	$\\
\smallskip
52	&8& $4.9418_{-0.0336}^{+0.0342} $& $	0.3605_{-0.0355}^{+0.0412} $& $	0.0172\pm0.0006 $& $6.85	$& $14.33	$\\
\smallskip
53	&8& $4.9418_{-0.0336}^{+0.0342} $& $	0.3605_{-0.0355}^{+0.0412} $& $	0.0172\pm0.0006 $& $6.85	$& $14.33	$\\
\hline\noalign{\smallskip} 
\end{tabular} 
\end{center}
\label{Tab:QPO}
\end{table*}

\begin{table*}
\caption{Parameters of the upper harmonic QPOs}
\begin{center}
\footnotesize
\begin{tabular}{llrrrrr}
\hline\noalign{\smallskip}
 \multicolumn{1}{c}{Snapshot} & \multicolumn{1}{c}{Obs.} & \multicolumn{1}{c}{$\nu_{\mr{0;QPO}}$ [Hz]}  & \multicolumn{1}{c}{$\Delta_{\mr{QPO}}$ [Hz] }  & \multicolumn{1}{c}{rms$_{\mr{QPO}}$ } &  \multicolumn{1}{c}{Q$_{\mr{QPO}}$}  & \multicolumn{1}{c}{$\sigma_{\mr{QPO}}$} \\
\hline\noalign{\smallskip}
1	&4& $6.6573_{-0.0873}^{+0.0851} $& $0.4470_{-0.0782}^{+0.0530} $& $0.0248_{-0.0018}^{+0.0017} $& $7.45	$& $6.89$\\
\smallskip
2	&4& $6.2603_{-0.0582}^{+0.0569} $& $0.1146_{-0.1146}^{+0.1123} $& $0.0192_{-0.0037}^{+0.0023} $& $27.31	$& $2.59$\\
\smallskip
3	&4& $6.7984_{-0.1103}^{+0.1014} $& $0.5406_{-0.1131}^{+0.1246} $& $0.0236_{-0.0019}^{+0.0023} $& $6.29	$& $6.21$\\
\smallskip
4	&4& $6.3205_{-0.0030}^{+0.0314} $& $0.0012_{-0.0012}^{+0.0750} $& $0.0152_{-0.0022}^{+0.0020} $& $42.14	$& $3.45$\\
\smallskip
5	&4& $6.8994_{-0.0520}^{+0.0446} $& $0.3657_{-0.0811}^{+0.0657} $& $0.0227_{-0.0017}^{+0.0015} $& $9.43	$& $6.68$\\
\smallskip
6	&4& $6.8254_{-0.0423}^{+0.0416} $& $0.3277_{-0.0745}^{+0.1091} $& $0.0232_{-0.0014}^{+0.0014} $& $10.41	$& $8.29$\\
\smallskip
7	&4& $7.3225_{-0.1204}^{+0.1291} $& $0.7145_{-0.1688}^{+0.2442} $& $0.0251\pm0.0024 $& $5.12	$& $5.23$\\
\smallskip
8	&4& $8.2624_{-0.2324}^{+0.1819} $& $0.3038_{-0.3038}^{+0.2990} $& $0.0181_{-0.0040}^{+0.0037} $& $13.60	$& $2.26$\\
\smallskip
9	&4& $8.2252_{-0.0627}^{+0.0571} $& $0.3771_{-0.0698}^{+0.0894} $& $0.0219_{-0.0013}^{+0.0014} $& $10.91	$& $8.42$\\
\smallskip
10	&4& $9.1198_{-0.1426}^{+0.1063} $& $0.5647_{-0.1824}^{+0.1892} $& $0.0210_{-0.0021}^{+0.0022} $& $8.07	$& $5.00$\\
\smallskip
11	&4& $9.3354_{-0.0103}^{+0.0079} $& $0.0010_{-0.0009}^{+0.0018} $& $0.0116\pm0.0011 $& $2593.17	$& $5.27$\\
\smallskip
12	&4& $8.4692_{-0.0005}^{+0.3366} $& $0.0011_{-0.0011}^{+0.5269} $& $0.0133_{-0.0015}^{+0.0026} $& $8.04	$& $4.43$\\
\smallskip
13	&4& $9.6192_{-0.1511}^{+0.2734} $& $0.2001_{-0.1106}^{+0.0982} $& $0.0145_{-0.0018}^{+0.0015} $& $24.04	$& $4.03$\\
\smallskip
14	&4& $11.0839_{-0.1411}^{+0.1345} $& $0.5803_{-0.2873}^{+0.3346} $& $0.0165_{-0.0019}^{+0.0017} $& $9.55	$& $4.34$\\
\smallskip
15	&5& $11.3310_{-0.1471}^{+0.0006} $& $0.0003_{-0.0003}^{+0.2292} $& $0.0114_{-0.0019}^{+0.0017} $& $24.72  $& $3.00$\\
\smallskip
16	&5& $11.3310	 	       $& $ 0.0003	              $& $0.0106_{-0.0018}^{+0.0015} $& $24.72	$& $2.94$\\
\hline\noalign{\smallskip} 
\end{tabular} 
\end{center}
\label{Tab:QPOuh}
\end{table*}

\begin{table*}
\caption{Parameters of additional features}
\begin{center}
\footnotesize
\begin{tabular}{llrrrrr}
\hline\noalign{\smallskip}
 \multicolumn{1}{c}{Snapshot} & \multicolumn{1}{c}{Obs.} & \multicolumn{1}{c}{$\nu_{\mr{0;QPO}}$ [Hz]}  & \multicolumn{1}{c}{$\Delta_{\mr{QPO}}$ [Hz] }  & \multicolumn{1}{c}{rms$_{\mr{QPO}}$ } &  \multicolumn{1}{c}{Q$_{\mr{QPO}}$}  & \multicolumn{1}{c}{$\sigma_{\mr{QPO}}$} \\
\hline\noalign{\smallskip}
1	&4& $	9.6455_{-0.4400}^{+0.6419} 	  $& $0.5050_{-0.5050}^{+0.8060} $& $0.0096_{-0.0043}^{+0.0042} $& $9.55	$& $1.12$\\
\smallskip
2	&4& $	7.5430_{-0.5178}^{+0.2184} $& $0.6379_{-0.5090}^{+0.4390} $& $0.0190_{-0.0051}^{+0.0053} $& $5.91        $& $1.86$\\
\smallskip
3	&4& $	1.7243_{-0.0050}^{+0.0002} $& $\le 0.0341	  $& $0.0075_{-0.0026}^{+0.0030} $& $25.28	$& $1.44$\\
\smallskip
4	&4& $	1.7242_{-0.0134}^{+0.0000}	  $& $0.0001_{-0.0001}^{+0.0311} $& $0.0121\pm0.0014 $& $27.70 	$& $4.32$\\
\smallskip
5	&4& $	9.2968_{-0.4165}^{+0.3248} $& $0.2865_{-0.2865}^{+1.1518} $& $0.0111_{-0.0037}^{+0.0045} $& $16.20 	$& $1.50$\\
\smallskip
10	&4& $	2.4611_{-0.0844}^{+0.0867} $& $0.7234_{-0.1004}^{+0.1740} $& $0.0278_{-0.0007}^{+6.0758} $& $1.70	$& $19.86$\\
\smallskip
14	&4& $	2.6830_{-0.1725}^{+0.1441} $& $0.4781_{-0.1922}^{+0.1987} $& $0.0179_{-0.0046}^{+0.0036} $& $2.81	$& $1.95$\\
\smallskip
18	&5&$	7.3404_{-0.3404}^{+0.3408} $& $\le 0.3830 	  $& $0.0053_{-0.0012}^{+0.0009} $& $9.58	$& $2.21$\\
\smallskip
22	&5&$	3.8148_{-0.3186}^{+0.0448} $& $0.0431_{-0.0431}^{+0.2582} $& $0.0057_{-0.0016}^{+0.0020} $& $44.26	$& $1.78$\\
\smallskip
23	&5&$	2.9982_{-0.2808}^{+0.0018} $& $1.1856_{-0.2633}^{+0.3081} $& $0.0230_{-0.0026}^{+0.0029} $& $1.26	$& $4.42$\\	
\smallskip
24	&5&$	3.5728_{-0.0785}^{+0.2877} $& $\le 0.0772 	  $& $0.0048_{-0.0010}^{+0.0008} $& $23.14	$& $2.4$0\\	
\smallskip
31	&6&$	1.3589_{-0.0642}^{+0.0754} $& $0.2157_{-0.0926}^{+0.1653} $& $0.0118_{-0.0026}^{+0.0030} $& $3.15	$& $2.27$\\
\smallskip
39	&7&$	2.5551_{-0.0659}^{+0.0678} $& $0.4074_{-0.1019}^{+0.1273} $& $0.0129_{-0.0013}^{+0.0014} $& $3.14	$& $4.96$\\
\smallskip
46b	&7&$	2.5990_{-0.0727}^{+0.0515} $& $	0.0575_{-0.0575}^{+2.7122} $& $	0.0061\pm0.0014 $& $	22.60	$& $2.18$\\
\hline\noalign{\smallskip} 
\end{tabular} 
\end{center}
\label{Tab:QPOaf}
\end{table*}

\clearpage
\section{Power density spectra for each snapshot}

\begin{figure*}
\resizebox{\hsize}{!}{\includegraphics[clip,angle=0]{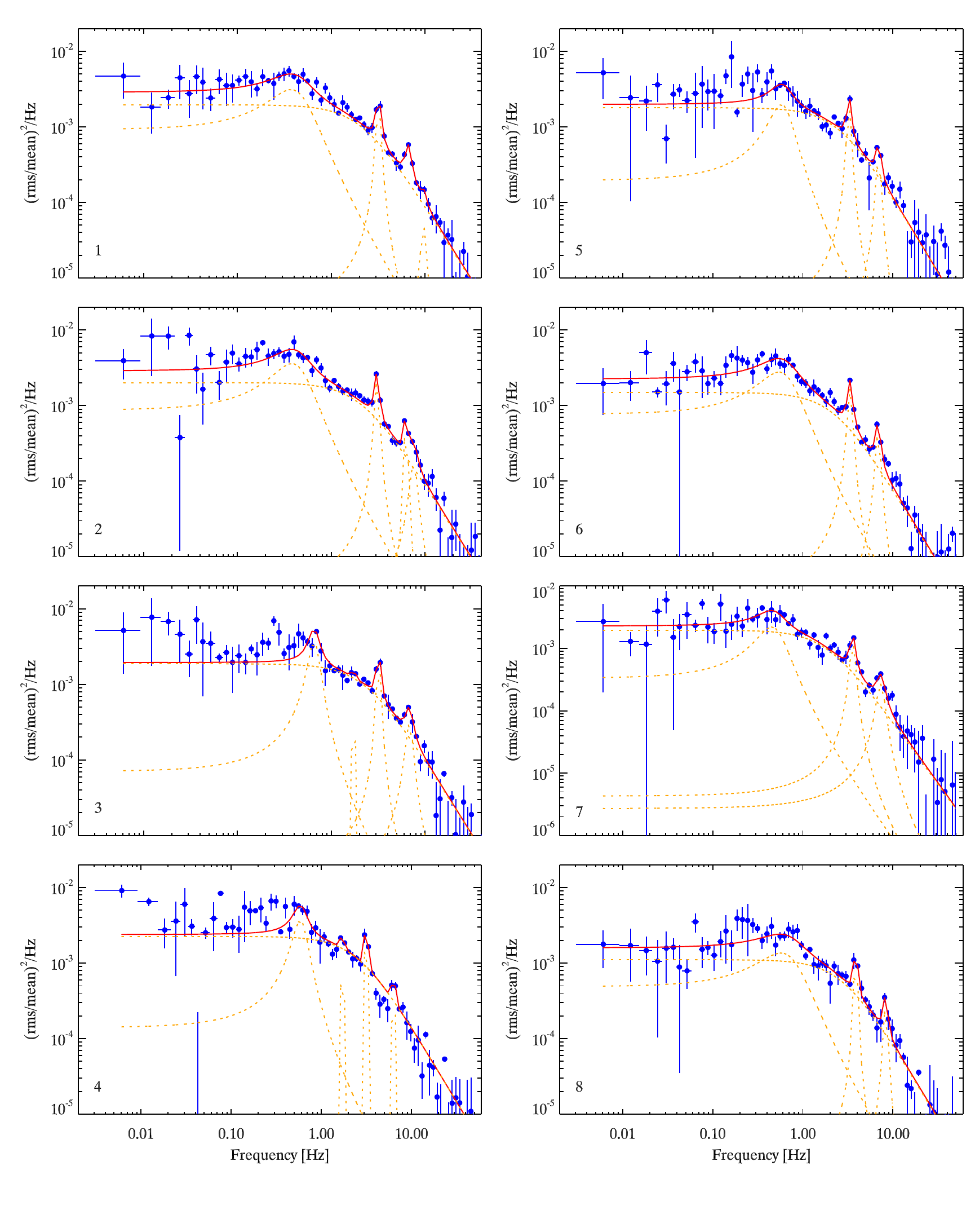}}
\caption{Power density spectra for each snapshot.}
\label{Fig:pds1}
\end{figure*}

\begin{figure*}
\resizebox{\hsize}{!}{\includegraphics[clip,angle=0]{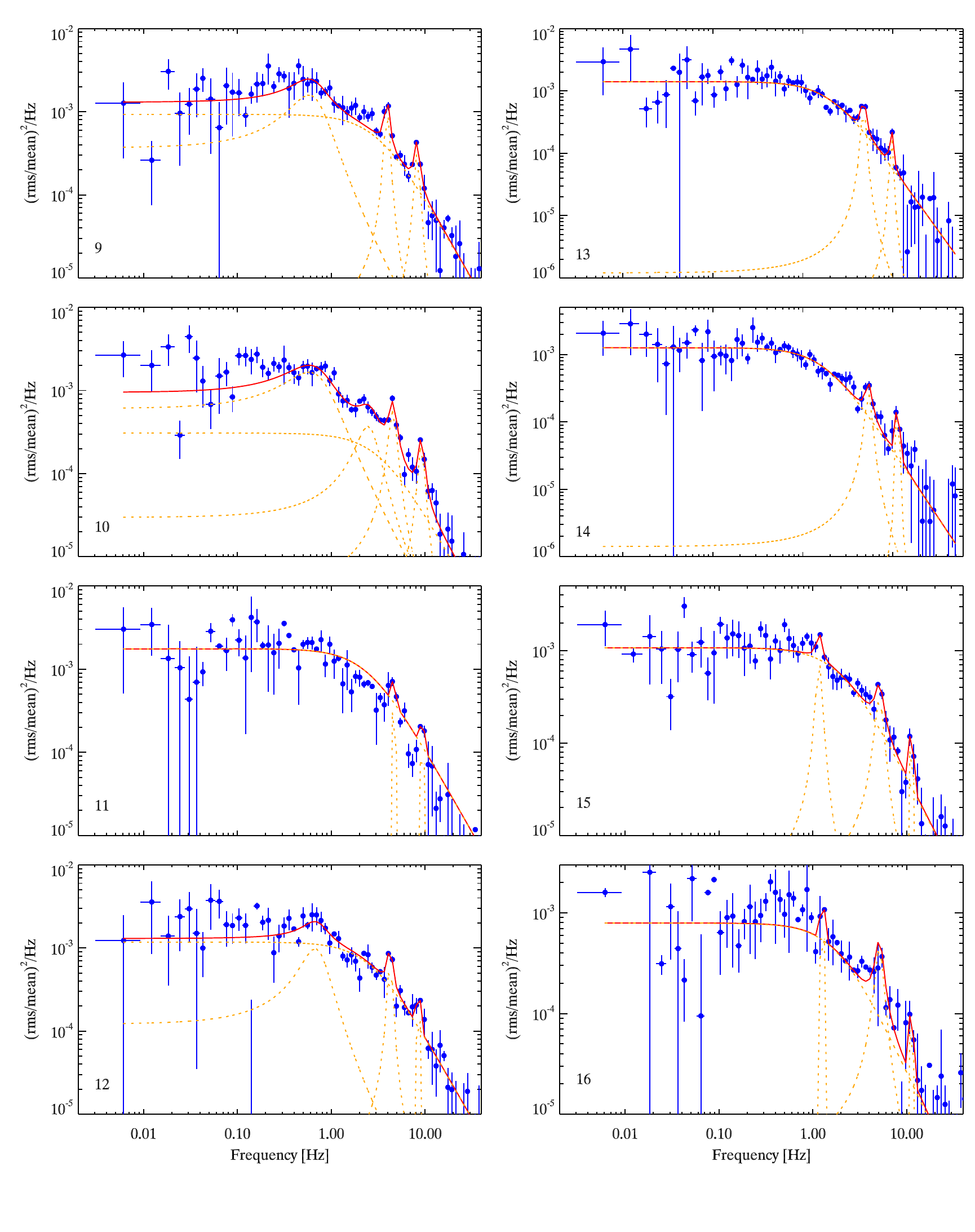}}
\addtocounter{figure}{-1}
\caption{continued.}
\label{Fig:pds2}
\end{figure*}

\begin{figure*}
\resizebox{\hsize}{!}{\includegraphics[clip,angle=0]{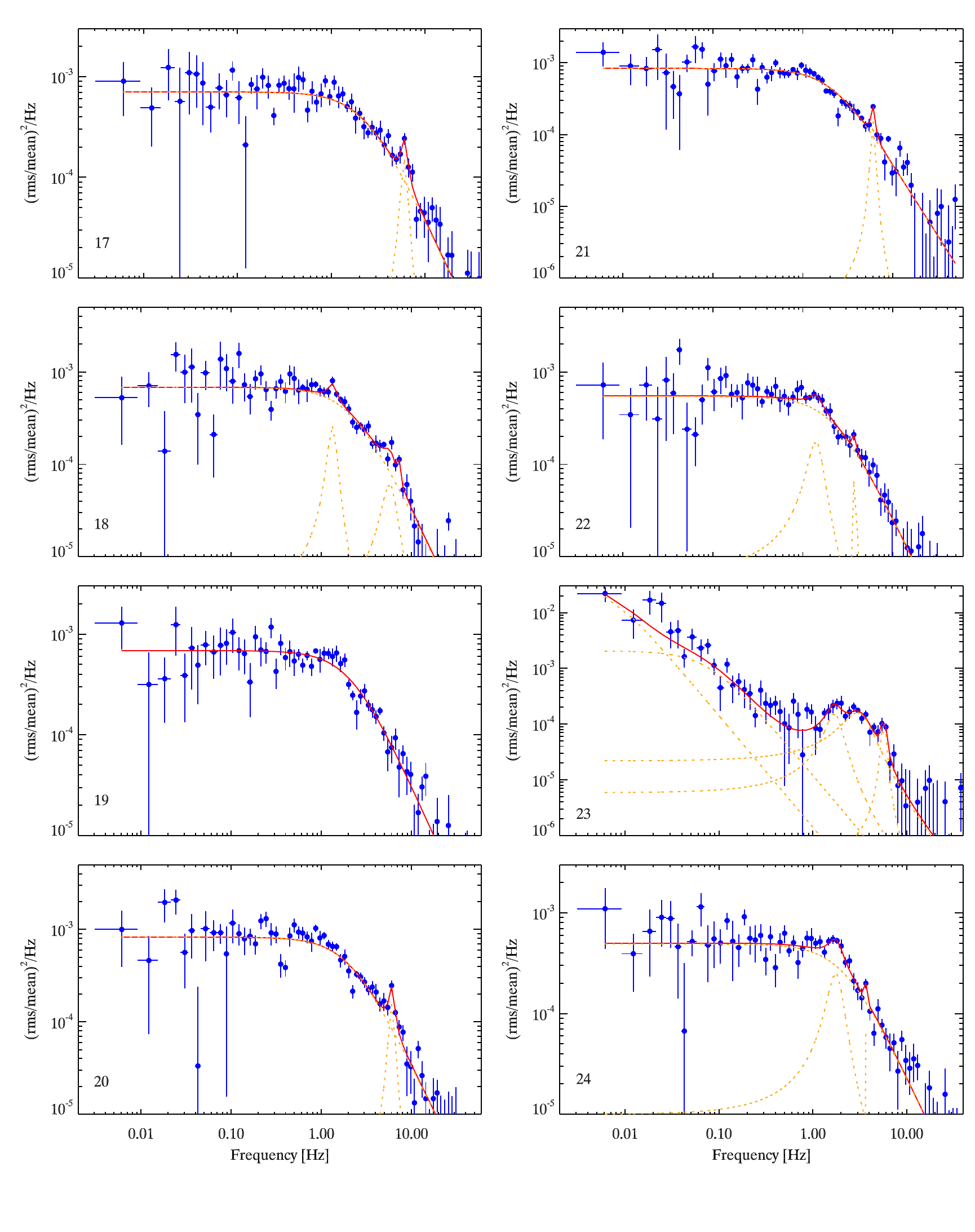}}
\addtocounter{figure}{-1}
\caption{continued.}
\label{Fig:pds3}
\end{figure*}

\begin{figure*}
\resizebox{\hsize}{!}{\includegraphics[clip,angle=0]{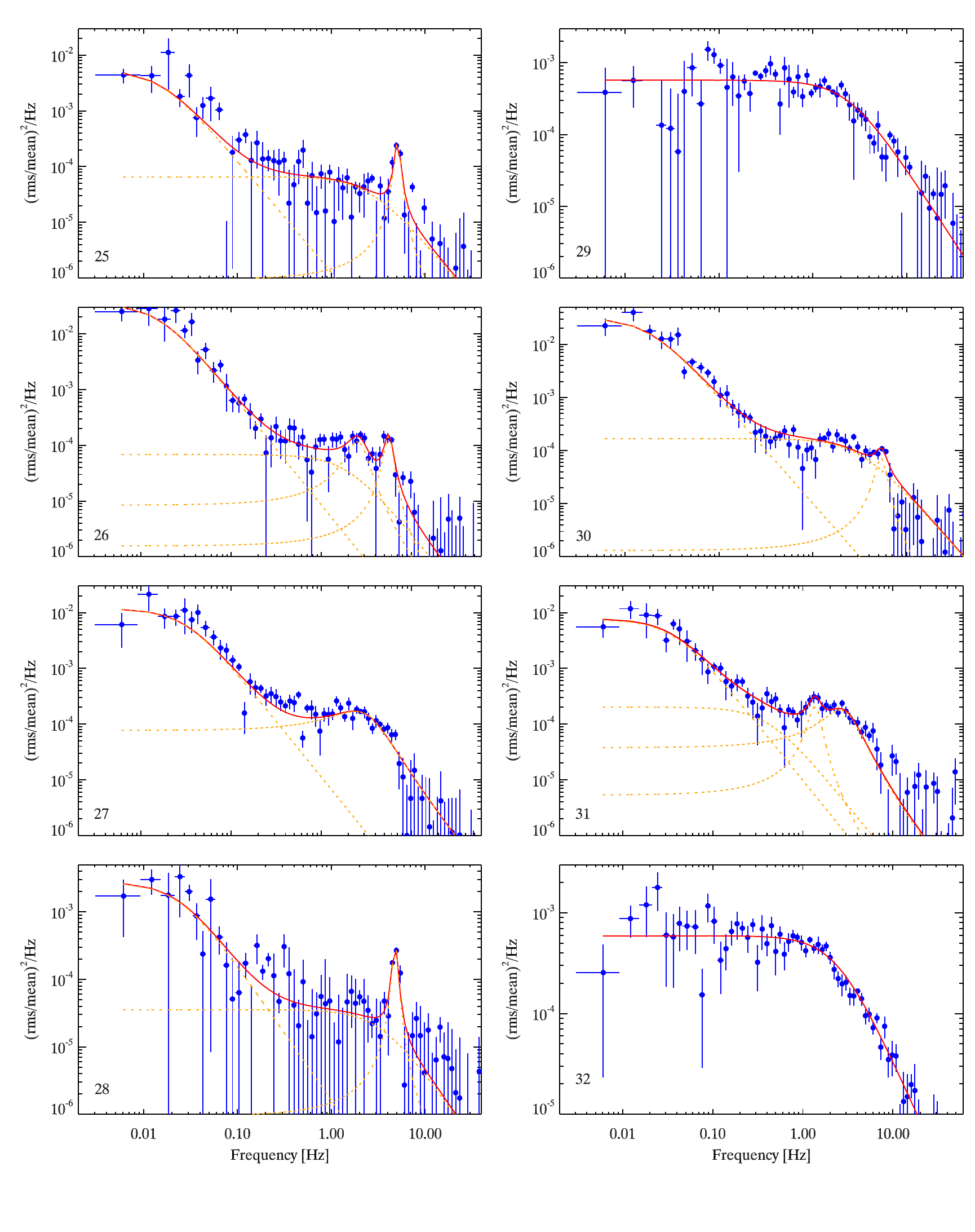}}
\addtocounter{figure}{-1}
\caption{continued.}
\label{Fig:pds4}
\end{figure*}

\begin{figure*}
\resizebox{\hsize}{!}{\includegraphics[clip,angle=0]{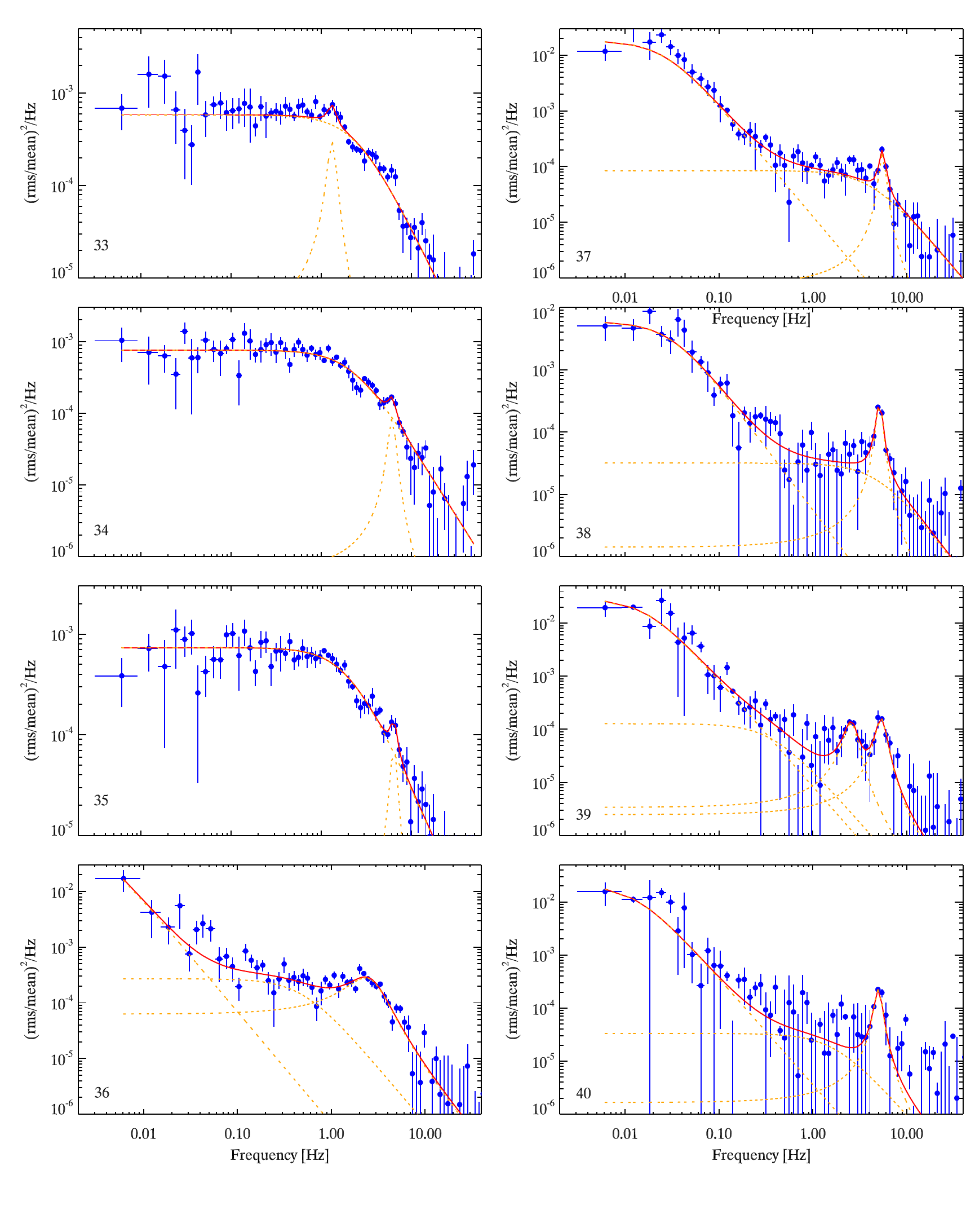}}
\addtocounter{figure}{-1}
\caption{continued.}
\label{Fig:pds5}
\end{figure*}

\begin{figure*}
\resizebox{\hsize}{!}{\includegraphics[clip,angle=0]{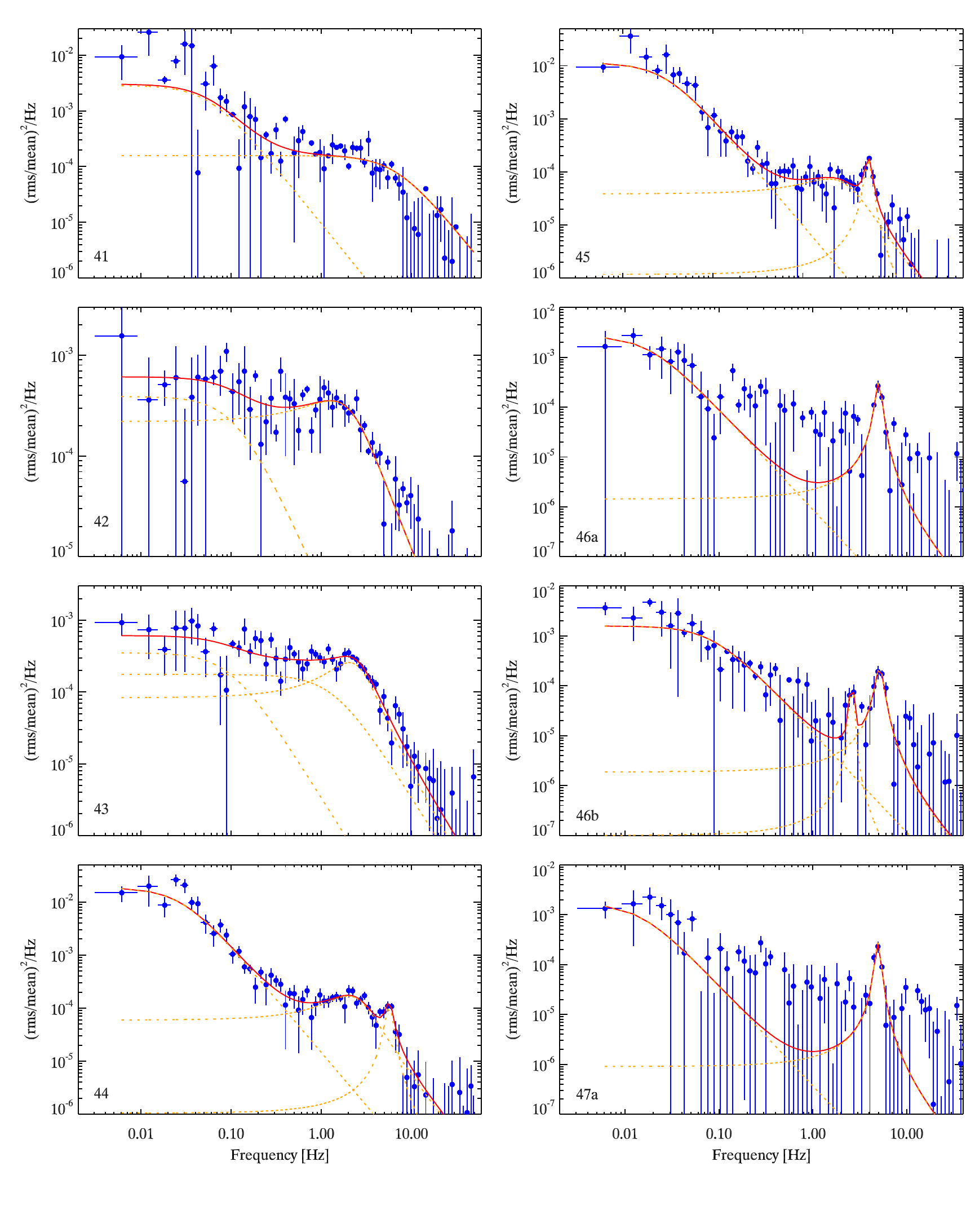}}
\addtocounter{figure}{-1}
\caption{continued.}
\label{Fig:pds6}
\end{figure*}

\begin{figure*}
\resizebox{\hsize}{!}{\includegraphics[clip,angle=0]{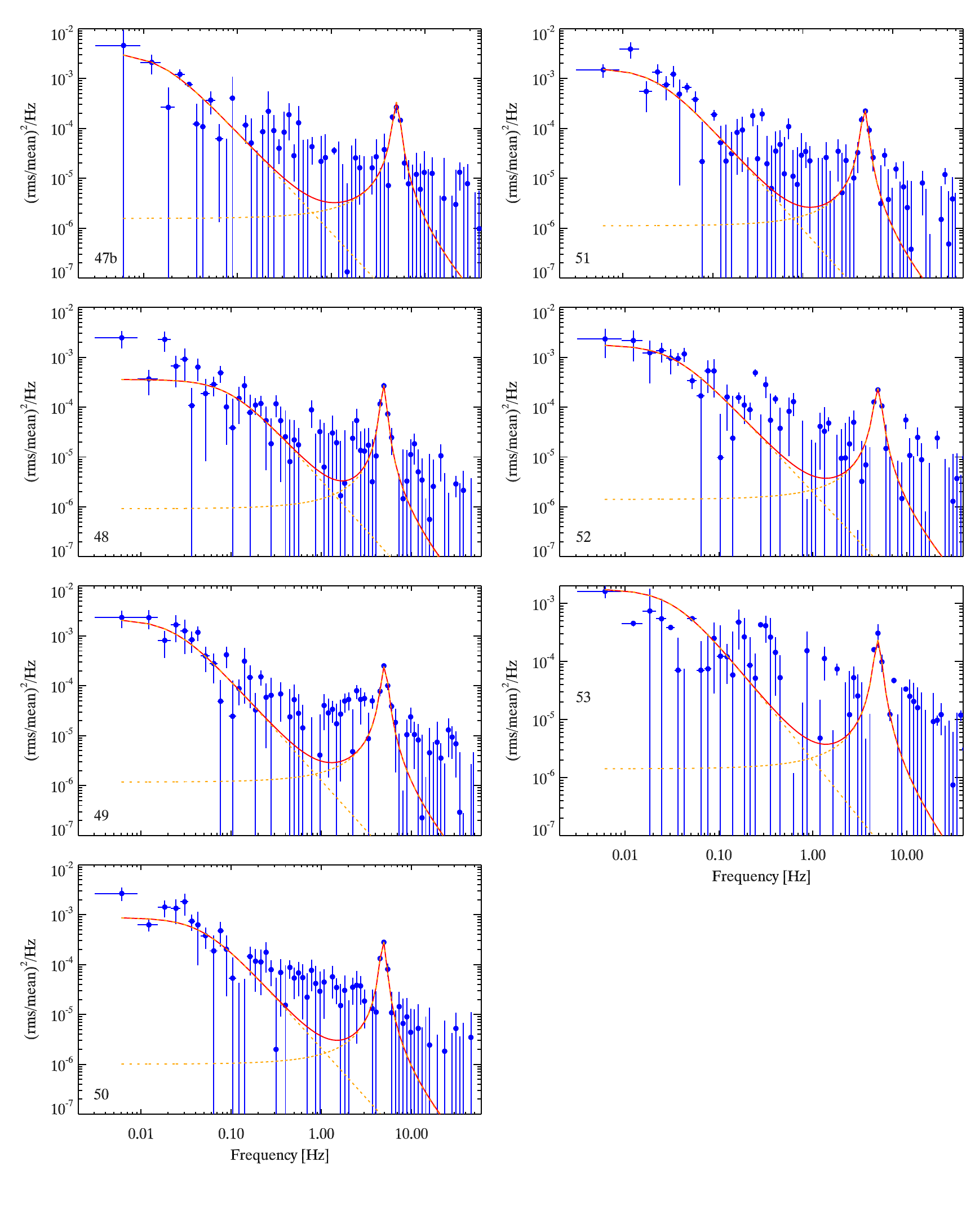}}
\addtocounter{figure}{-1}
\caption{continued.}
\label{Fig:pds7}
\end{figure*}


\bsp	
\label{lastpage}
\end{document}